\documentclass{revtex4}

\usepackage{graphicx}
\usepackage{amssymb}
\usepackage{amsmath}
\usepackage{simplemargins}
\usepackage{color}

\setleftmargin{1cm}
\setrightmargin{1cm}
\settopmargin{1cm}
\setbottommargin{1cm}

\newcommand{\iainnote}[1]{#1}

\begin{document}

\title{Closed-form stochastic solutions for non-equilibrium dynamics and inheritance of cellular components over many cell divisions}

\author{Iain G. Johnston and Nick S. Jones}

\begin{abstract}
Stochastic dynamics govern many important processes in cellular biology, and an underlying theoretical approach describing these dynamics is desirable to address a wealth of questions in biology and medicine. Mathematical tools exist for treating several important examples of these stochastic processes, most notably gene expression, and random partitioning at single cell divisions or after a steady state has been reached. Comparatively little work exists exploring different and specific ways that repeated cell divisions can lead to stochastic inheritance of unequilibrated cellular populations. Here we introduce a mathematical formalism to describe cellular agents that are subject to random creation, replication, and/or degradation, and are inherited according to a range of random dynamics at cell divisions. We obtain closed-form generating functions describing systems at any time after any number of cell divisions for binomial partitioning and divisions provoking a deterministic or random, subtractive or additive change in copy number, and show that these solutions agree exactly with stochastic simulation. We apply this general formalism to several example problems involving the dynamics of mitochondrial DNA (mtDNA) during development and organismal lifetimes.
\end{abstract}

\maketitle

\section{Introduction} 


Stochastic dynamics underlie a multitude of processes in cellular biology \cite{mcadams1997stochastic, altschuler2010cellular, elowitz2002stochastic, kaern2005stochasticity, raj2008nature, tsimring2014noise}. Understanding the sources of this randomness within and between cells is a central current challenge in quantitative biology \cite{johnston2012chaos}.  Noise has been found to affect processes including stem cell fate decisions \cite{chang2008transcriptome, clayton2007single}, bet-hedging in bacterial phenotypes \cite{fraser2009chance, kussell2005bacterial}, cancer development \cite{brock2009non}, and responses to apoptosis-inducing factors \cite{bastiaens2009systems, spencer2009non}, illustrating the fact that a theoretical understanding of stochastic cellular biology is of great importance in medical and biological problems.

Partitioning of cellular components at cell divisions provides a considerable source of stochasticity in cell biology \cite{huh2010nongenetic}. Huh and Paulsson have shown that uneven segregration of cellular constituents at mitosis can contribute significantly to cell-to-cell differences in levels of cellular components and proteins in a population, focusing on stochasticity in protein inheritance between sister cells \cite{huh2011random}. In addition to variability in protein levels, there is evidence that random partitioning of mitochondria at cell divisions can lead to substantial extrinsic variability in the physical and chemical attributes, and behavioural phenomena including differentiation propensity, in a population of cells \cite{johnston2012mitochondrial}. 

Historically, mathematical modelling, including the use of birth-and-death processes, have provided a theoretical foundation with which to describe phenomena in stochastic biology \cite{novozhilov2006biological}. Early work on the problem of the stochastic evolution of cellular constituents in a population of dividing cells was performed in the context of protein levels in bacterial cells by Berg \cite{berg1978model} and Rigney \cite{rigney1979stochastic}, who made analytic progress with birth-and-death models coupled to cell division in the case of a steady-state population of cells with constant birth and death rates. \iainnote{A famous example of stochastic analysis of cellular systems is the Luria-Delbr\"{u}ck treatment of population statistics of bacterial mutations \cite{luria1943mutations}, which has been addressed by several analyses including generating function approaches \cite{zheng1999progress}}. Matrix equations, with operators corresponding to the processes of birth, death, and partitioning, have also been used to obtain numerical results on stochastic effects in populations of dividing cells \cite{jones1980stochastic, marshall2007stability}, and a framework of nonequilibrium statistical mechanics has been used to derive general properties of protein content of dividing cells \cite{brenner2007nonequilibrium}. Stochastic models examining the behaviour of a continuous variable (for example, protein concentrations) have also been widely used in cellular biology \cite{grima2008modelling}.

Several stochastic models have been formulated for systems involving compartmentalised elements which replicate and are partitioned as compartments divide. An early example of this approach is due to Dowman \cite{dowman1973implications}, which considers mean copy number and extinction probability of cellular elements for specific birth, death, and partitioning dynamics. Other applications are in the study of growing intestinal crypts, in which a collection of crypts, each containing a replicating quantity of stem cells, grow and divide according to a branching process \cite{loeffler1991stochastic}, and in the stochastic corrector model, whereby a population of replicators with given concentrations of constituents divide and propagate \cite{grey1995reexamination}. More recently, Swain \emph{et al.} \cite{swain2002intrinsic} have derived analytic results describing stochastic gene expression in dividing cells after a steady limit cycle has been reached. Quantitative results, in terms of integrals over kinetic rate dynamics, have been obtained for stochastic gene expression where cellular components are binomially partitioned at cell divisions \cite{rausenberger2008quantifying}. The variance resulting from more general partitioning dynamics of cellular components has also been addressed for single cell divisions \cite{huh2010nongenetic,huh2011random}.

In this article, we focus on birth-immigration-death (BID) dynamics rather than the well-studied model dynamics of stochastic gene expression. \iainnote{With this framework we attempt to provide a model for the stochastic dynamics of cellular components other than gene products; we particularly focus on mitochondrial DNA (mtDNA) in several examples.} Populations of hundreds or thousands of mtDNA molecules are typically present in eukaryotic cells, replicating and degrading somewhat independently of the cell cycle. MtDNA encodes vital bioenergetic machinery, making it an important target for stochastic analysis. We will consider a variety of partitioning regimes and, where possible, an arbitrary number of cell divisions, and aim to derive closed-form generating functions describing the dynamics of our model cellular components. In so doing, we avoid assumptions about equilibrium behaviour and restrictions to lower-order moments of copy number distributions, aiming to produce a non-equilibrium theory to describe stochastic dynamics in full distributional detail, at arbitrary times during any cell cycle. Our specific consideration of mtDNA dynamics provides a theoretical framework with which a class of models, often analysed numerically through simulation, can be descibed analytically.

In the first section we introduce our formalism and derive generating functions for BID dynamics with binomial partitioning and the inheritance of a deterministically or randomly reduced or increased complement of the parent cell's population. We next illustrate the exact agreement between our theory and stochastic simulation, and \iainnote{investigate several example biological questions, regarding the dynamics of mtDNA during a copy number bottleneck and over organismal lifetimes involving many cell divisions. We further harness the power afforded by a generating function approach to explore extinction probabilities in these systems,} and also perform our analysis for a given number of cell divisions in which cellular components are deterministically partitioned, and randomly partitioned in clusters. Finally, we discuss the implications of our mathematical formalism for approaches to stochastic biology.


\section{Model and analysis} 

In this section we detail the approach we take to obtain generating functions describing the stochastic dynamics of cellular agents subject to birth-immigration-death dynamics and stochastic partitioning at cell divisions. We first illustrate how the generating function describing agent dynamics within a cell cycle is derived. We next consider how this solution may be extended over cell divisions, using an assumption (later shown to be true) about the functional form of the expression describing the inheritance of agents at cell divisions. We then show that this extended solution gives rise to an overall generating function containing factors that are the solutions to recursion relations, where each recursive step corresponds to a cell division. We obtain solutions for these recursion relations, thus yielding general generating functions for dynamics over arbitrarily many cell divisions. Finally, we validate our early assumption for several important specific cases of inheritance dynamics and derive the specific solutions in these case.

\subsection{Agents within a cell cycle}

\begin{figure}
\includegraphics[width=18cm]{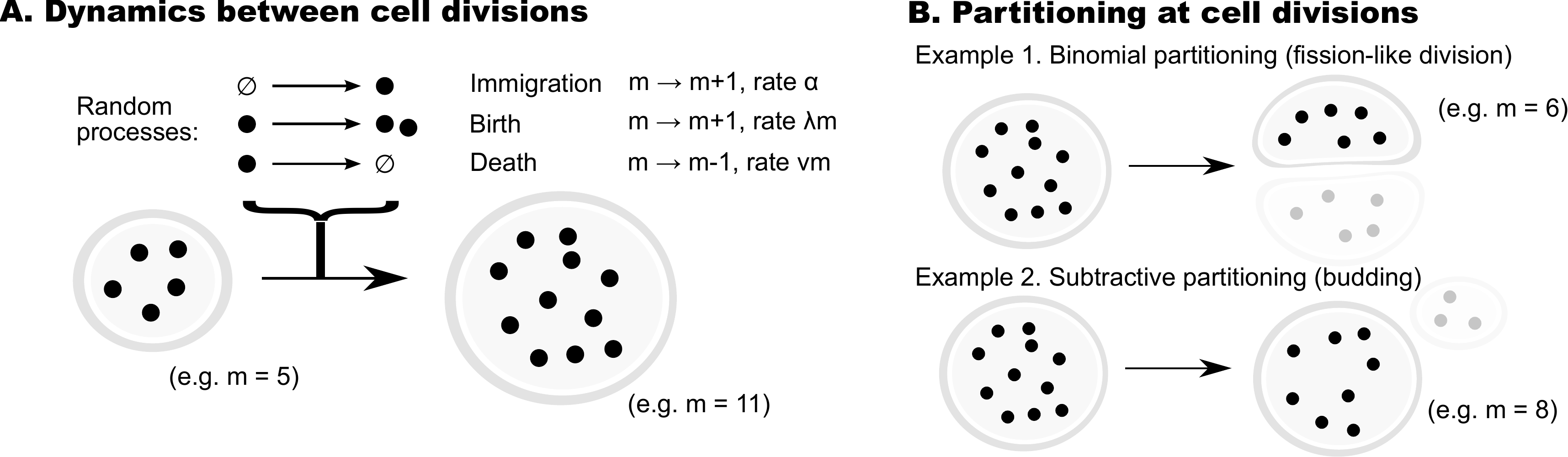}
\caption{\iainnote{Illustration of birth-immigration-death dynamics with random partitioning. We will derive expressions for the statistics of copy number $m$ of cellular agents over successive cell generations, separated by divisions. (A) Between cell divisions, agents may be produced, replicated, or degraded; each is a Poissonian event. The copy number $m$ of agents in a cell is a random variable that changes with these dynamics. (B) At cell divisions, the copy number of agents changes according to a different type of random event. Two possibilities are illustrated here: the binomial partitioning of agents into two daughter cells (one of which will be tracked in the next generation), and the loss of a random number of agents to a smaller bud (the larger remaining cell will be tracked in the next generation).}}
\label{illus}
\end{figure}

We consider birth-immigration-death (BID) dynamics, where agents are created ($\emptyset \rightarrow \bullet$) with rate $\alpha$, replicate ($\bullet \rightarrow \bullet \bullet$) with rate $\lambda$, and are removed ($\bullet \rightarrow \emptyset$) with rate $\nu$ (see Fig. \ref{illus}A). These processes are referred to as immigration, birth, and death respectively, and are assumed to be Poissonian, with non-time-varying rates (though see Results). We will later consider setting some of these parameters to zero, as special cases of the overall BID dynamics. \iainnote{We note that some literature refers to our `immigration' term (producing agents from nothing) as `birth', using the term `replication' to describe the production of agents from existing agents; the specific symbols used to denote these rates also vary. For compatibility with a wide body of literature we adopt the nomenclature above.} 

The dynamics of populations under BID dynamics are given by the corresponding master equation, describing the time evolution of the probability $P(m,t)$ that the system contains $m$ agents at time $t$:

\begin{equation}
\frac{d P(m, t)}{dt} = \alpha P(m-1,t) + \nu (m+1) P(m+1, t) + \lambda (m-1) P(m-1, t) - (\alpha + \nu m + \lambda m) P(m, t), \label{mastereqn1} 
\end{equation}

with initial condition

\begin{equation}
P(m,0) = \delta_{m\,m_0}. \label{mastereqn2}
\end{equation}

We are concerned with the generating function $G(z,t) = \sum_{m=0}^{\infty} z^m P(m, t)$ for the distribution of cellular components in a general set of conditions. Once this generating function has been calculated, then, by its definition, all information about the distribution $P(m, t)$ can be obtained through taking its derivatives, for example, $\mathbb{E}(m) = \left. \frac{\partial G}{\partial z} \right|_{z=1}$, $\mathbb{V}(m) = \left. \left(\frac{\partial^2 G}{\partial z^2} + \frac{\partial G}{\partial z}  - \left(\frac{\partial G}{\partial z} \right)^2 \right) \right|_{z=1}$, $P(m)  =  \left. \frac{1}{m!} \frac{\partial^m G}{\partial z^m} \right|_{z=0}$. The generating function corresponding to Eqns. \ref{mastereqn1}-\ref{mastereqn2} obeys

\begin{eqnarray}
\frac{\partial G(z, t)}{\partial t} & = &  \alpha (z-1) G(z,t) + ( \nu (1-z) + \lambda(z^2 - z) )  \frac{\partial G(z, t)}{\partial z} \label{gfeqn1} \\
G(z,0) & = & z^{m_0}, 
\end{eqnarray}

\iainnote{the solution to which is well known} (see Appendix):

\begin{eqnarray}
G(z,t) & = & \underbrace{ \left( \frac{ \nu - \lambda }{\lambda e^{(\lambda-\nu)t} (z-1) - \lambda z + \nu} \right)^{\frac{\alpha}{\lambda}} }_{\xi(z,t)}  \left. \underbrace{ \left( \frac{\nu e^{(\lambda-\nu)t} (z-1) - \lambda z + \nu}{\lambda e^{(\lambda-\nu)t} (z-1) - \lambda z + \nu} \right) }_{g(z,t)} \right.^{m_0} \label{underbraces} \\
& \equiv & \xi(z,t) (g(z,t))^{m_0} \label{gfsolutionmaintext}
\end{eqnarray}

where the final line, using the underbraced substitutions in Eqn. \ref{underbraces}, casts the solution in a form that will be useful later.

\subsection{Partitioning of agents at cell divisions}

We now consider a series of cell divisions, linked by quiescent periods governed by the within-cell-cycle dynamics above. Time is accounted for throughout this model in the following manner. \emph{A full cell cycle is assumed to take time $\tau$, after which a division occurs. A variable $t$ measures the elapsed time since the most recent cell division.} As divisions occur every interval $\tau$, $t$ is always less than or equal to $\tau$. If $n$ divisions have occurred, the total elapsed time since the initial state is $n \tau + t$. We will here assume that $\tau$ is constant, showing later that this assumption can be relaxed when our approach is extended to deal with many different dynamic phases.

We will generally write $P_i(m,t|m_0)$ for the probability of observing $m$ agents at a time $t$ after the $i$th cell division, given initial condition $m_0$ (at the start of the cell cycle before the first cell division). Hence, $P_0(m,t|m_0)$ is the probability of observing copy number $m$ at time $t$, given that zero cell divisions have occurred; and $P_{i-1}(m, \tau | m_0)$ is the probability of observing $m$ agents at a time $\tau$ after the $(i-1)$th cell division (i.e. immediately before the $i$th cell division). 

Consider the $i$th cell division in a series of divisions. \emph{Throughout this article, we will use subscript $a$ to denote `after' and subscript $b$ to denote `before' a cell division:}, thus, we write $m_{i,b}$ for copy number before the division and $m_{i,a}$ for copy number afterwards. We assume that $m_{i,b} \geq m_{i,a}$ always. We write $P_{\delta}(a | b)$ for the probability of observing $a$ agents after a cell division, given $b$ agents before that division. The generating function at time $t$ after the cell division is then given by

\begin{eqnarray}
&& G_i(z, t) \nonumber \\
 & = & \sum_m \sum_{m_{i,b} = 0}^{\infty} \sum_{m_{i,a} = 0}^{m_{i,b}} z^m P_0(m, t | m_{i,a}) P_{\delta}(m_{i,a} | m_{i,b}) P_{i-1}(m_{i,b}, \tau | m_0) \\ 
& = & \sum_{m_{i,b} = 0}^{\infty} \sum_{m_{i,a} = 0}^{m_{i,b}} G(z, t | m_{i,a}) P_{\delta}(m_{i,a} | m_{i,b}) P_{i-1}(m_{i,b}, \tau | m_0) \\ 
& = & \sum_{m_{i,b} = 0}^{\infty} \sum_{m_{i,a} = 0}^{m_{i,b}} \xi(z,t) g(z,t)^{m_{i,a}} P_{\delta}(m_{i,a} | m_{i,b}) P_{i-1}(m_{i,b}, \tau | m_0) \label{finalppterm} 
\end{eqnarray}

We will make the assumption that the expression in Eqn. \ref{finalppterm} may be reduced to the form

\begin{eqnarray}
G_i(z,t) & \equiv & \xi(z,t) \phi(z,t)  \sum_{m_{i,b} = 0}^{\infty} \theta(g(z,t))^{m_{i,b}} P_{i-1}(m_{i,b}, \tau | m_0) \label{assumptionline1}
\end{eqnarray}

where $\theta$ and $\phi$ are functions to be determined, given knowledge of a particular partitioning rule. The partitioning mechanisms that we will subsequently consider can all be cast in this form, as we will demonstrate.

We now consider the overall generating function describing a set of cell divisions. To represent the sum over all possible copy numbers before and after all cell divisions between divisions $j$ and $i$, we introduce the notation

\begin{equation}
\sum'_{i, j} \equiv \sum_{m_{i,b} = 0}^{\infty} \sum_{m_{i,a} = 0}^{m_{i,b}} \sum_{m_{i-1,b} = 0}^{\infty} \sum_{m_{i-1,a} = 0}^{m_{i-1,b}}  ... \sum_{m_{j,b} = 0}^{\infty} \sum_{m_{j,a} = 0}^{m_{j,b}}.
\end{equation}

This combination of sums takes into account all possible states before and after each cell division $i, i-1, ..., j$, for $i \geq j$. We note that the ordering of sums here progresses backwards in time from left to right: the leftmost two sums sum over all configurations related to the most recent cell division $i$, the next two sum over all configurations related to the preceding cell division $i-1$, and so on. The final probability distribution is then

\begin{equation}
P_n(m, t | m_0) = \sum'_{n,1} P_0(m, t | m_{n,a}) \prod_{i = 1}^{n-1} \Phi_i,
\end{equation}

where $\Phi_i$ is a `probability propagator' of the form

\begin{equation}
\Phi_i = P_{\delta}(m_{i,a} | m_{i,b}) P_0(m_{i,b}, \tau | m_{i-1, a}),
\end{equation}

representing the probability that a cell, which started with $m_{i-1,a}$ units after division $i-1$, grew to have $m_{i,b}$ units, of which $m_{i,a}$ units were inherited by the next daughter cell after division $i$. The chain of divisions can be terminated at $n$ divisions in the past by setting $m_{0,a} \equiv m_0$ as the initial condition of the ancestor cell. Thus, a subscript $1$ labels the first cell division, and a subscript $n$ labels the most recent of $n$ cell divisions.

The overall generating function after $n$ divisions is

\begin{eqnarray}
&& G_n(z, t | m_0) \nonumber \\
& = & \sum_m  \sum'_{n,1} z^m P_0(m, t | m_{n,a}) \prod_{i = 1}^{n-1} \Phi_i \\
& = & \sum'_{n,1} G(z, t | m_{n,a}) \prod_{i = 1}^{n-1} \Phi_i \label{induct1} \\
& = &  \sum'_{n-1, 1} \sum_{m_{n,b} = 0}^{\infty} \sum_{m_{n,a} = 0}^{m_{n,b}}  \xi(z,t)  g(z,t)^{m_{n, a}} P_{\delta}(m_{n,a} | m_{n,b}) P_0(m_{n,b}, \tau | m_{n-1, a}) \prod_{i = 1}^{n-2} \Phi_i 
\end{eqnarray}

Generalising the approach of Rausenberger \& Kollmann \cite{rausenberger2008quantifying}, we now employ the assumption in Eqn. \ref{assumptionline1} to write

\begin{eqnarray}
&& \sum'_{n-1, 1} \sum_{m_{n,b} = 0}^{\infty} \sum_{m_{n,a} = 0}^{m_{n,b}}  \xi(z,t)  g(z,t)^{m_{n, a}} P_{\delta}(m_{n,a} | m_{n,b}) P_0(m_{n,b}, \tau | m_{n-1, a}) \prod_{i = 1}^{n-2} \Phi_i \\
& = & \xi(z,t) \phi(z,t) \sum'_{n-2,1} \sum_{m_{n,b} = 0}^{\infty} \theta(z,t)^{m_{n,b}} P_0(m_{n,b}, \tau | m_{n-1,a}) \prod_{i = 1}^{n-2} \Phi_i \label{assumptionline2} \\
& = & \xi(z,t) \phi(z,t) \sum'_{n-2,1} G(\theta(g(z,t)), \tau | m_{n-1,a}) \prod_{i = 1}^{n-2} \Phi_i \\
& \equiv & \xi(z,t) \phi(z,t) \sum'_{n-2,1} G(z_1, \tau | m_{n-1,a}) \prod_{i = 1}^{n-2} \Phi_i, \label{induct2}
\end{eqnarray}

where in the Eqn. \ref{induct2} we have changed variables $z_1 \equiv \theta(g(z,t))$. Comparing Eqns. \ref{induct1} and \ref{induct2}, we can see that this process can be followed inductively. Each further step through the induction corresponds to another cell division, extracts a prefactor of $\phi(z_i, \tau)$, and causes another change of variables $z_{i+1} = \theta( g(z_i, \tau) )$. Extending this induction to $n$ cell divisions, the overall generating function after $n$ divisions is then

\begin{equation}
\boxed{ G_n(z,t) = \left( \prod_{i=1}^n \xi(z_i, \tau) \right) \left( \prod_{i=1}^n \phi(z_i, \tau) \right) \xi(z,t) h_0(z,t)^{m_0} } \label{generalgf}
\end{equation}

where $h_i$ and $z_i$ are the solutions to

\begin{eqnarray}
h_i(z,t) & = & g \left( \theta( h_{i+1}) , \tau  \right) \,;\, h_n(z,t) = g (z, t) \label{mainrecur1} \\
z_{i+1} & = & \theta(g(z_i, \tau)) \,;\, z_1 = \theta(g(z, t)). \label{mainrecur2}
\end{eqnarray}

The recurrence relations Eqns. \ref{mainrecur1}-\ref{mainrecur2} are rather similar, but we retain their distinction for mathematical convenience, also noting that their indexing runs in opposite directions through time. Hence, the boundary condition for $h_i$ arises from the most recent cell division, corresponding to $i = n$; the boundary condition for $z_i$ also arises from the most recent division, but in this indexing this corresponds to $i=1$.

It can be noted that for $n=0$, the products in Eqn. \ref{generalgf} vanish and the boundary condition $h_n \equiv h_0 = g(z,t)$ leads to $G_0(z,t) = \xi(z,t) g(z,t)^{m_0}$ as required.

For BID dynamics, the product and final term in Eqn. \ref{generalgf} are analytically tractable for several important inheritance regimes, allowing us to write the generating function in an exact form. \iainnote{We will first analyse two partitioning regimes of importance for biological modelling, and illustrate how this approach produces statistics of interest for mtDNA populations under these regimes. We will later explore other partitioning regimes.}

\subsection{Binomial inheritance}

We first consider the case where agents are partitioned binomially at cell divisions. In this case, the following identities hold:

\begin{eqnarray}
P_{\delta}(m_{i,a} | m_{i,b}) & = & \binom{m_{i,b}}{m_{i,a}} 2^{-m_{i,b}}; \\
\sum_{m_{i,a} = 0}^{m_{i,b}}  & \xi(z,t) & g(z,t)^{m_{i, a}} P_{\delta}(m_{i,a} | m_{i,b}) \nonumber \\
 & = &  \xi(z,t) \left( \frac{1}{2} + \frac{g(z,t)}{2} \right)^{m_{i,b}}; \label{dumb1} \\
\text{and so}\,\, \phi(z,t) & = & 1; \label{dumb2} \\
\theta(g(z,t)) & = & \left( \frac{1}{2} + \frac{g(z,t)}{2} \right), \label{dumb3}
\end{eqnarray}

where Eqns. \ref{dumb2}-\ref{dumb3} follow by comparing Eqn. \ref{dumb1} with Eqn. \ref{assumptionline1}. We are thus concerned with the solutions to the recurrence relations

\begin{eqnarray}
z_i & = & \frac{1}{2} + \frac{g(z_{i-1}, \tau)}{2} \,;\, z_1 = \frac{1}{2} + \frac{g(z,t)}{2}. \\
h_i & = & g \left( \frac{1}{2} + \frac{h_{i+1}}{2}, \tau \right) \,;\, h_n = g (z, t). \label{recur2}
\end{eqnarray}

We will introduce the symbols $l \equiv e^{(\lambda-\nu) \tau}$ and $l' \equiv e^{(\lambda-\nu)t}$ for convenience here and throughout. In the Appendix we solve these related systems of equations, showing that the solutions take the form


\begin{eqnarray}
h_i & = & \frac{\kappa_{11} 2^i + \kappa_{12} l^i}{\kappa_{13} 2^i + \kappa_{14} l^i}, \label{binomialhmaintext} \\
z_i & = & \frac{\kappa_{21} 2^i + \kappa_{22} l^i}{\kappa_{23} 2^i + \kappa_{24} l^i}
\end{eqnarray}


with $\kappa_{11} = l^n l'(z-1) (\lambda + \nu (l-2)), \kappa_{12} = \kappa_{14} = 2^n (\lambda (l' - z(l+l'-2)) + \nu(l-2)), \kappa_{13} = l^n l' (z-1) \lambda (l-1), \kappa_{21} = \kappa_{23} = - l (\lambda l' (\-1) + (l-2) (\lambda z - \nu)), \kappa_{22} = l'(z-1)(l(\lambda+\nu)-2\nu), \kappa_{24} = 2 \lambda l' (z-1)(l-1)$.



We also show in the Appendix that the first product in Eqn. \ref{generalgf} takes the form 

\begin{equation}
\left( \frac{B_1^{n+1} B_2^{-n-1} (A_2 + B_2) (-A_1/B_1; \rho_A / \rho_B )_{n+1}} {(A_1+B_1) (-A_2/B_2; \rho_A / \rho_B )_{n+1}} \right)^{\gamma}, \label{firstproductform}
\end{equation}

where $A_1 = 2 \lambda l' (l-1) (z-1), A_2 = \lambda l' (z-1) (e^{(\lambda - \nu)\tau} (l-2) + l), B_1 = B_2 = l(\lambda l' (1-z) - (l-1) (\lambda z - \nu)), \rho_A = l, \rho_B = 2,$ and $\gamma = \alpha / \lambda$. $(a; q)_n$ is the $q$-Pochhammer symbol defined by $(a;q)_n \equiv \prod_{k=0}^{n-1} (1 - aq^k)$. \iainnote{While this symbol is hard to interpret intuitively, we will see that expressions for important moments of distributions often only involve particular derivatives of the symbol that reduce to simple algebraic expressions (see Results). In addition, this term vanishes in the $\alpha = 0$ case where immigration dynamics can be ignored.}

Combining $h_0$ from Eqn. \ref{binomialhmaintext} and Eqn. \ref{firstproductform}, we therefore use Eqn. \ref{generalgf} to yield a closed-form generating function for BID dynamics with binomial partitioning at cell divisions (the full form is explicitly presented in the Appendix). In the Results section we will demonstrate the efficacy of this generating function solution (and subsequent solutions) by showing that moments derived from the generating function exactly match stochastic simulation (Fig. \ref{figproof}). This algebraic generating function for an arbitrary number of cell divisions extends a previous solution presented in integral form from Ref. \cite{rausenberger2008quantifying}, with the advantage that the methodology allows straightforward analytic investigation of this and a wider class of systems; we next demonstrate this generalisation with a new inheritance regime which we will demonstrate has biological applicability.


\subsection{Random additive or subtractive inheritance}
We now consider the case where a number of agents are lost or gained at each cell division, and this number is itself a random variable \iainnote{(we will consider the case where this number is a fixed constant later)}. For mathematical convenience we shall first assume that a certain number of agents are lost at partitioning, and that this number is taken from a binomial distribution with population size $2 \eta$ and probability $\frac{1}{2}$, so that the average loss number is $\eta$ (it is straightforward to see that a negative value for the $\eta$ parameter corresponds to the gain of a number of elements identically distributed). In this case, by considering the possible values of $n$, the number of agents lost at a division, we obtain


\begin{eqnarray}
P_{\delta}(m_{i,a} | m_{i,b}) & = & \sum_{n=0}^{2 \eta} \delta_{m_{i,a}, m_{i,b} - n} \binom{2 \eta}{n} 2^{-2 \eta}; \\
 \sum_{m_{i,a} = 0}^{m_{i,b}} & \xi(z,t) & g(z,t)^{m_{i, a}} P_{\delta}(m_{i,a} | m_{i,b}) \nonumber \\
& = &  \xi(z,t) \left( \frac{1}{2} + \frac{1}{2 g(z,t)} \right)^{2 \eta} g(z,t)^{m_{i,b}}; \\
\text{and so}\,\, \phi(z,t) & = &  \left( \frac{1}{2} + \frac{1}{2 g(z,t)} \right)^{2 \eta}; \label{phicase2} \\
\theta(g(z,t)) & = & g(z,t).
\end{eqnarray}

The solutions to the corresponding recurrence relations are derived in the Appendix and are:

\begin{eqnarray}
h_i & = & \frac{l^n l' \nu (z-1) + l^i (\nu - \lambda z)} {l^n l' \lambda (z-1) + l^i (\nu - \lambda z)}; \\
z_i & = & \frac{l^i l' \nu (z-1) + l (\nu - \lambda z)}{l^i l' \lambda (z-1) + l (\nu - \lambda z)}.
\end{eqnarray}

The first product in Eqn. \ref{generalgf} is $\prod_{i=1}^n \xi(z_i, \tau)$, which we show in the Appendix takes the form of Eqn. \ref{firstproductform} with $A_1 = \lambda l' (z-1), A_2 = \lambda l' e^{(\lambda - \nu) \tau} (z-1), B_1 = B_2 = l(\nu - \lambda z), \rho_A = l, \rho_B = 1, \gamma = \alpha / \lambda$. The second product term is $\prod_{i=1}^n (1/2 + 1/(2 g(z_i,\tau)))^{2 \eta}$. We here introduce the symbols $x_1 \equiv \lambda (l^{-1} - 1) / (\lambda - \nu), x_2 \equiv \lambda (l - 1) / (\lambda - \nu)$. In the Appendix we show that this product takes the form of Eqn. \ref{firstproductform} with $A_1 = l' (\lambda + \nu) (-x_2)^n (z-1), B_1 = B_2 = -2 x_1^n (\lambda z - \nu), A_2 = 2 l' \nu (-x_2)^n (z-1), \rho_A = x_1, \rho_B = (-x_2), \gamma = 2 \eta$.

We then have a closed-form generating function for random subtractive partitioning and BID dynamics (the full form is presented in the Appendix). \iainnote{We note here that applying this approach to the case of loss at cell divisions does not explicitly restrict copy number to be non-negative, and care must therefore be taken in its application. If the dynamics under investigation are such that the probability of copy number $m < \eta$ is negligible, the absence of this restriction will have negligible influence on results extracted from the analysis. If low copy numbers are likely, this approach can still yield useful results if $\alpha = 0$, if expressions for $P(m,t)$ are derived and $P(m \leq 0, t)$ is treated as equivalent to $P(m = 0, t)$. This approximation is valid because the birth and death operations have $m=0$ as an absorbing state, so a copy number below zero can never subsequently exceed $m=0$. However, the simple expressions for $\mathbb{E}(m, t)$ and $\mathbb{V}(m,t)$ in terms of generating function derivatives will yield incorrect results in these cases. The approximation will fail in cases where a non-negligible probability of attaining zero copy number is coupled with dynamics involving immigration (as opposed to birth). In this case, a specific boundary rule must be written in Eqn. \ref{mastereqn1}; we have been unable to find closed-form solutions in this case.}

\section{Results and applications}

\subsection{Comparison with stochastic simulation}

\begin{figure}
\includegraphics[width=15cm]{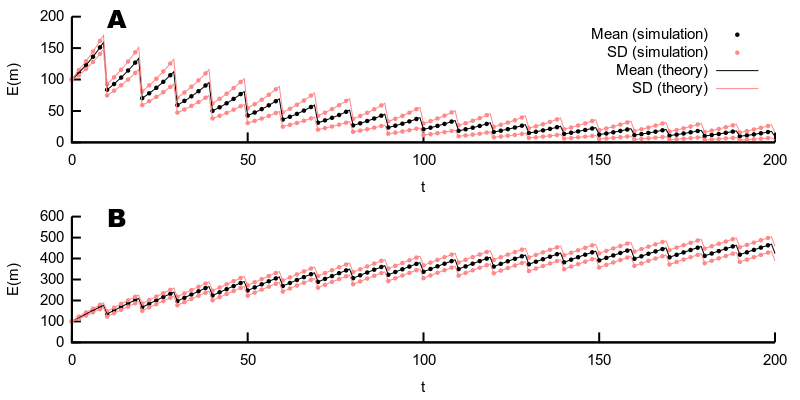}
\caption{\textbf{Comparison of analytic results for first two moments with stochastic simulation.} Trajectories of copy number mean and standard deviation resulting from our analytic results (lines) and stochastic simulation (points). (A) Birth-immigration-death (BID) dynamics within cell cycles; binomial partitioning at cell divisions. (B) BID dynamics within cell cycles; loss of a binomially-distributed number of agents ($p=\frac{1}{2}, N = 100$) at each cell division. Other parameters for these systems are given in the text. Theory and simulation match exactly in all cases.}
\label{figproof}
\end{figure}

In Fig. \ref{figproof} we compare the analytic results for copy number mean and variance, derived from our generating functions above, with the results obtained over $10^5$ simulations using Gillespie's stochastic simulation algorithm \cite{gillespie1977exact}. In order to compute moments arising from the generating functions for subtractive inheritance it is necessary to compute a small number of values corresponding to derivatives of the $q$-Pochhammer symbol $(a;q)_n$ with respect to the first argument $a$. We are not aware of a general analytic form for this derivative, but these values can be evaluated to arbitrary precision by symbolic software or through numerical perturbation, by evaluating $( (a; q)_n - (a+\epsilon; q)_n ) / \epsilon$ for suitably small $\epsilon$. In this case we take `suitably small' to mean `yielding a estimate converged to the required degree of accuracy'. In addition, in several cases which arise (for example, $a = 0$, emerging from our analysis below), this perturbative approach yields an analytic solution. In the Supplementary Information we provide Mathematica notebooks illustrating these calculations (and other calculations in this article).

We choose arbitrary parameterisations for these confirmatory studies: we choose $m_0 = 100, \tau = 10$, for binomial partitioning we use $\alpha = 0.2, \lambda  = 0.06, \nu = 0.01$, and for subtractive inheritance we use $\alpha = 10, \lambda = 0.01, \nu = 0.02$, with $\eta = 50$ for deterministic and random subtraction. Analytic results exactly match simulation results throughout.

We proceed by considering two examples motivated by specific biological questions involving cellular populations of mitochondrial DNA (mtDNA).

\subsection{MtDNA bottlenecking: birth dynamics, binomial partitioning, changing population size}

In the case of the birth-death model with binomial partitioning, we can explore the levels of cellular noise introduced by controlled variation of the population size of cellular components without making continuous or steady-state approximations.

We will consider a number $r_{max}$ of dynamic phases labelled by $r$, where the rates $\lambda, \nu$ are constant within a phase but may take different values in different phases. To extend the above reasoning to describe different phases of dynamics it is necessary to compute the function $g_r(z, t)$ for each regime $r$, where $g_r(z, t)$ is the generating function using the appropriate parameters $\lambda_r, \nu_r$ for phase $r$ and calculated at $n = n_r$, the number of cell cycles of phase $r$. For consistency with the above approach, we label phases starting from a zero index, so the first phase corresponds to $r = 0$, and we use $r_{max}$ to denote the label of the final phase. Then we use

\begin{eqnarray}
h_{r_{max}} & = & g_{r_{max}}(z, t) \label{phaseseqn1} \\
h_{r} & = & g_r (h_{r+1}, 0) \\
G_{overall} & = & h_{0}^{m_0}, \label{phaseseqn2}
\end{eqnarray}

using induction over the different phases in the way way we used induction over different cell cycles above. Here we consider the changeover between phases by using the generating function at the start of the incoming phase. 

The system Eqns. \ref{phaseseqn1}-\ref{phaseseqn2} can be solved for arbitrarily many phases with different kinetic parameters, producing closed-form results for a wide range of different dynamic trajectories, including arbitrarily varying population size and cell doubling times (see Appendix). We illustrate this approach with the following simple two-phase model system. Initially, $m_0$ agents exist in a cell, with no associated cell-to-cell variability. These agents subsequently follow birth-only dynamics and binomial partitioning at cell divisions. The rate of birth is initially $\lambda_1$ in the first dynamic phase, changing to $\lambda_2$ after $n_1$ cell divisions. We choose $\lambda_2 = 2 \ln 2 / \tau - \lambda_1$, to ensure that mean copy number returns to $m_0$ after a further $n_1$ cell divisions. We will use $\lambda_2 \geq \lambda_1 \leq \ln 2 / \tau$, so the mean copy number either remains constant or initially drops to a minimum (the `bottleneck') before recovering.

For simplicity, we have here assumed that $\tau$, cell cycle length, is the same constant in each dynamic phase. This assumption can readily be relaxed by using different $\tau_i$, so that cell cycle length is labelled by the current dynamic phase. In this way, our formalism can be used to explore the dynamics of systems with arbitrarily varying (though deterministic) cell cycle lengths.

Using Eqn. \ref{generalgf} with Eqns. \ref{binomialhmaintext} and \ref{firstproductform}, the solution to Eqns. \ref{phaseseqn1}-\ref{phaseseqn2} for two dynamic phases is simply $G = g_1(g_2(z,t), 0)^{m_0}$, with $\alpha = 0$ and $\nu = 0$. After some algebra, we obtain, after $n_1$ cell divisions in the first phase and another $n_1$ in the second,

\begin{equation}
\mathbb{V}(m) = \frac{m_0 e^{-\lambda_1 n_1 \tau} (e^{\lambda_1 \tau} +2)(e^{\lambda_1 n_1 \tau} - 2^{n_1}) }{e^{\lambda_1 \tau} - 2}. \label{bottleneckvar}
\end{equation}

The minimum copy number attained immediately follows cell division $n_1$ and is thus of size $b = 2^{-n_1} m_0 e^{n_1 \lambda_1 \tau}$. This allows us to write the parameter $\lambda_1$ in terms of the bottleneck it produces, $\lambda_1 = \ln (2^{n_1} b / m_0 ) / ( n_1 \tau )$. Inserting this expression into Eqn. \ref{bottleneckvar} gives

\begin{equation}
\mathbb{V}(m) = \frac{m_0 (b - m_0)}{b} \left( \frac{(b/m_0)^{1/n_1} + 1}{(b/m_0)^{1/n_1} - 1} \right). \label{bottleneckvarb}
\end{equation}

Mitochondrial DNA (mtDNA) is observed to be present in fertilised oocytes at copy numbers around $10^5$ \cite{cree2008reduction, cao2007mitochondrial, wai2008mitochondrial}. During subsequent development, a pronounced copy number decrease occurs, as cells divide rapidly with little replication of mtDNA. The copy number per cell falls to a low bottleneck then recovers during later development. This mechanism is believed to ameliorate the inheritance of mutated mtDNA by increasing the cell-to-cell variability of mutant load in a cellular population and hence allowing cell-level selection to discard those cells that drift towards high mutant content. Substantial debate surrounds this topic \cite{wallace2013mitochondrial, carling2011implications}: competing mechanisms have been proposed to increase mutant load variability \cite{wai2008mitochondrial, cao2007mitochondrial}, and the size of the copy number bottleneck, its power to generate variance, and thus its biological importance have been questioned \cite{cao2007mitochondrial}.

Results from classical population genetics \cite{wright1942statistical, kimura1955solution} have been applied to the statistics of mtDNA populations \cite{wonnapinij2008distribution}, but even with refinements modelling fluctuations in the size of, and substructure in, the mtDNA population \cite{iizuka2001effective, maruyama1980genetic}, these results lack straightforward physical interpretability and the ability to address population statistics at arbitrary, non-steady-state points in developmental dynamics. The stochastic formalism we present has recently been used to address these issues, specifically in modelling the mtDNA bottleneck in mice \cite{usrecent}. Here we explore a more general question: what increase in copy number variance is possible through enforcing an mtDNA bottleneck of a specific size, and how does it relate to the size of that bottleneck? Results on the copy number statistics of mtDNA can then readily be extended to explore statistics of the mutant load of mtDNA, by considering two decoupled populations of mutant and wildtype mtDNA.
 
\begin{figure}
\includegraphics[width=15cm]{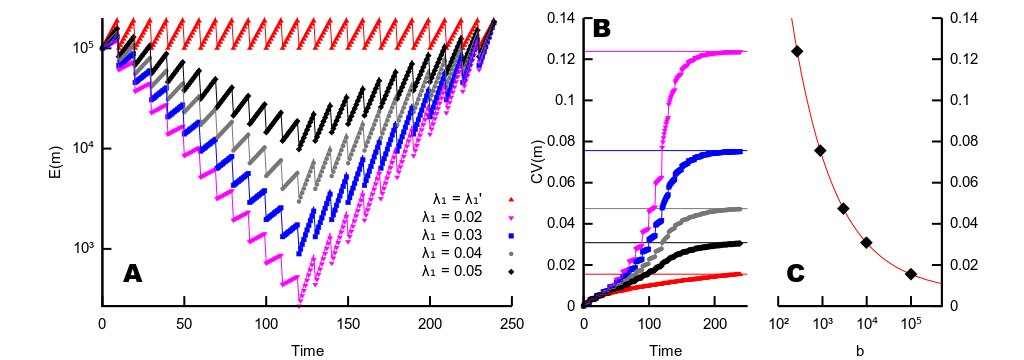} 
\caption{\textbf{Modelling copy number variability due to mtDNA bottlenecking.} (A) Trajectories of mean copy number of cellular agents born with rate $\lambda_1$ ($t < 120$) and $\lambda_2 = 2 \ln 2 / \tau - \lambda_1$ ($t \geq 120$). Lower $\lambda_1$ values enforce a smaller copy number bottleneck, with copy number recovering to its initial value after the bottleneck. $\lambda_1' = \ln 2 \tau$ is the value required to maintain constant mean copy number between cell divisions. Lines are analytic results; points are stochastic simulations. (B) Trajectories of coefficient of variation (CV) as different bottleneck sizes are imposed on the system. Horizontal lines give the analytic predictions for final CV derived from Eqn. \ref{bottleneckvarb}. Other lines are analytic results; points are stochastic simulations. (C) CV as a function of bottleneck size from Eqn. \ref{bottleneckvarb}; points show specific instances of stochastic simulation.}
\label{figbneck}
\end{figure}

In Fig. \ref{figbneck} we simulate the above system for $n_1 = 12$, $m_0 = 10^5$, roughly matching the above mtDNA copy number magnitudes and number of divisions \cite{lawson1994clonal} observed in the mouse germ line. We use various $\lambda_1$ values, reporting mean copy number and copy number coefficient of variation (CV). It can be observed that lower $\lambda_1$ rates lead to more pronounced copy number bottlenecks, resulting in increased CVs that match the predictions derived from the above analysis. 

Eqn. \ref{bottleneckvarb} thus analytically describes the increased variance due to a copy number bottleneck under specific circumstances, whereby a copy number $m_0$ is decreased to a minimum $b$ over $n_1$ cell divisions then raised to its original value over a further $n_1$ cell divisions. Lower bottleneck sizes $b$ lead to exponential increases in the cell-to-cell variability associated with an mtDNA population. The assumptions within our illustrative model here can also straightforwardly be relaxed and closed-form solutions derived for more general dynamics, \iainnote{and a closed-form expression for the probability distribution function can also be derived using the above approach.} \iainnote{We note than in this example, the variance does not converge to a final fixed value: longer times will result in higher variances. This feature can be altered using a model involving more homeostatic dynamics (see next subsection) or, in biology, may conceivably be dealt with on a cellular level by retaining cells with certain copy number statistics.}


\subsection{Relaxed replication of mtDNA: immigration-death dynamics and various inheritance regimes}

A quantitative model for mtDNA dynamics throughout organismal lifetimes has been proposed to account for the intuitive feature that mtDNA copy number should be subject to cellular control \cite{chinnery1999relaxed}. This `relaxed replication' model has influenced a wide range of studies on mtDNA dynamics in many contexts from human disease \cite{dimauro2001mitochondrial} to forensics \cite{melton2004mitochondrial}; its quantitative behaviour has been explored (considering low-order moments without cell divisions) in the contexts of nuclear control on mtDNA \cite{capps2003model} and through simulation studies in topics including ageing mtDNA \cite{elson2001random}, the effect of anti-retroviral drugs on mitochondrial \cite{payne2011mitochondrial}, and many others. As variability in mtDNA can have important physiological consequences \cite{wallace2013mitochondrial}, we aim here to analytically extend this model beyond a mean-only treatment to a more general (and realistic) situation both explicitly modelling the stochastic dynamics of individual mtDNA production and degradation and including different forms of cell division dynamics.

The governing dynamics of copy number $m$ in the model are


\begin{equation} 
\frac{dm}{dt} = \left\{ \begin{array}{ll} 
\tilde{\alpha} m_{opt} / \tilde{\tau} - \tilde{\alpha} m / \tilde{\tau} & \mbox{if $m \leq \frac{\tilde{\alpha} m_{opt}}{\tilde{\alpha} - 1}$;} \\
-m / \tilde{\tau} & \mbox{otherwise.} \end{array} \right. 
\label{eqncapps}
\end{equation}

where $m_{opt}$ is a `target' copy number, $\tilde{\tau}$ is the timescale of mtDNA degradation, and $\tilde{\alpha} > 1$ is a parameter of the model describing nuclear feedback (in the original papers, $\tilde{\alpha}$ and $\tilde{\tau}$ are respectively assigned the symbols $\alpha$ and $\tau$: we use tildes to avoid ambiguity with our analysis). For $\tilde{\alpha}$ not much greater than 1 and an initial condition $m_0 < m_{opt}$, the probability of $m > \tilde{\alpha} m_{opt} / (\tilde{\alpha} - 1)$ is very low; we will thus assume that the contribution of the term in Eqn. \ref{eqncapps} corresponding to $m > \tilde{\alpha} m_{opt} / (\tilde{\alpha} - 1)$ is negligible.

When considering cell divisions in the above model, the meaning of $m_{opt}$ needs to be made explicit. We will consider $m_{opt}$ to be the target copy number for the end of a cell cycle, immediately before division. We then note that the first term in Eqn. \ref{eqncapps} contains a combination of a immigration term (independent of copy number) and a death term (dependent on copy number); we can therefore write the model as Eqn. \ref{mastereqn1} with $\alpha \equiv \beta m_{opt}$, $\lambda = 0$, and $\nu \equiv \beta$, defining the new parameter $\beta \equiv \tilde{\alpha} / \tilde{\tau}$. We can then use the above treatment to obtain the generating function of the relaxed replication model under different partitioning regimes (full forms given in the Appendix) and moments of interest of the copy number distribution in each case.



\begin{figure*}
\includegraphics[width=18cm]{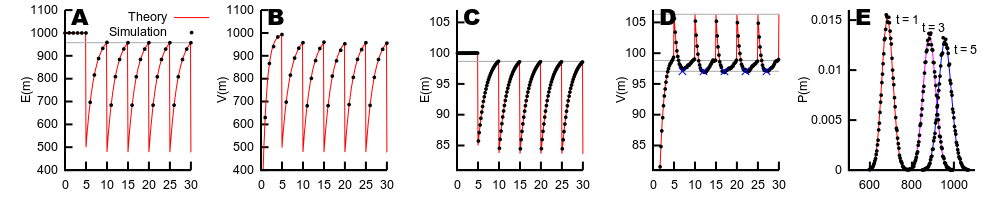}
\caption{Copy number statistics for the relaxed replication model of mtDNA. Lines are analytic results; points are from stochastic simulation. (A) $\mathbb{E}(m)$ and (B) $\mathbb{V}(m)$ for binomial partitioning, modelling dividing cells, with $m_0 = m_{opt} = 1000$, $\beta = \frac{1}{2}$. $\mathbb{V}(m)$ and $\mathbb{E}(m)$ converge to the same trajectory; grey line gives analytic result for $\mathbb{E}(m)$ at the end of a cell cycle. (C) $\mathbb{E}(m)$ and (D) $\mathbb{V}(m)$ for random subtractive partitioning, modelling budding cells, with $m_0 = 100 = m_{opt} = 100$, $\beta = \frac{1}{2}$, $\eta = 15$. Grey lines give analytic results for $\mathbb{E}(m)$ at the end of a cell cycle, and $\mathbb{V}(m)$ at start, end, and minimum point (blue crosses) in the cell cycle. (E) Comparison of theory and stochastic simulation in illustrative snapshots of probability distribution functions at given times in the binomial case using Eqn. \ref{relrepsoln}.} 
\label{relaxed}
\end{figure*}

\textbf{Binomial partitioning.} First we consider the case where mtDNA molecules are binomially partitioned at cell divisions. The dynamics in this case are illustrated in Fig. \ref{relaxed} for $m_{opt} = m_0 = 1000$, $\tilde{\alpha} = 5$, $\tilde{\tau} = 10$, and cell cycle length $\tau = 5$. This parameter set was chosen for compatibility with original work on the relaxed replication model: mtDNA copy numbers around 1000 are biological reasonable, $\tilde{\alpha} = 5$ is an intermediate value of the nuclear feedback parameter explored in Ref. \cite{capps2003model}, and $\tilde{\tau} = 10$ corresponds to an mtDNA half-life of $10 \ln 2 \simeq 7$, compatible with the range (in days) of half-lives assumed in Ref. \cite{capps2003model}. The cell cycle length $\tau = 5$ was chosen both for rough biological applicability (corresponding to cells dividing every 5 days) and to illustrate transient behaviour of the model. 

It can be observed that the variance and the mean converge on the same behaviour; the difference between the two is straightforwardly found to be $\mathbb{E}(m) - \mathbb{V}(m) = 4^{-n} m_0 e^{-2 \beta(t + n \tau)}$, clearly decreasing to zero with cell divisions $n$. The mean copy number immediately before a division $\mathbb{E}(m, \tau)$ takes a value that approaches, but does not reach, $m_{opt}$. Some algebra (see Appendix) shows that this value, in the long time limit, is 

\begin{equation}
\mathbb{E}(m, \tau) = m_{opt} - \frac{1}{2 e^{\beta \tau} - 1} m_{opt}.
\end{equation}

The generating function analysis also allows an expression to be derived for the probability distribution function for mtDNA copy number (see Appendix). 

%

\begin{equation}
P(m,t) = (1/m!) (-b)^m (1-b)^{m_0 - m} e^{-a}\, U \left(-m, 1-m+m_0, a(b-1)/b \right), \label{relrepsoln}
\end{equation}

where $U(.,.,.)$ is the confluent hypergeometric function, $a = m_{opt} ( 1 - e^{-\beta t} + (2^{-n} e^{-\beta t} (e^{\beta \tau} - 1)(2^n - e^{-\beta n \tau} ))/(2 e^{\beta \tau} - 1))$ and $b = 2^{-n} e^{-\beta (t + n \tau)}$. As $n \rightarrow \infty$, this converges to the simpler expression

\begin{equation}
P(m,t) = \frac{1}{m!} \left( \frac{m_{opt} (1 - 2e^{\beta \tau} + e^{-\beta(t - \tau)} )}{2 e^{\beta \tau} - 1} \right)^m \exp \left( \frac{m_{opt} (1 - 2e^{\beta \tau} + e^{-\beta(t - \tau)} )}{2 e^{\beta \tau} - 1} \right).
\end{equation}


Eqn. \ref{relrepsoln} thus provides a complete solution for the relaxed replication model with binomial cell divisions (see Fig. \ref{relaxed} for a comparison with stochastic simulation for $\mathbb{E}(m)$, $\mathbb{V}(m)$ and $P(m)$; other statistics and quantities of interest are readily extracted from the generating function).

\textbf{Subtractive partitioning.} Next we consider the case of dynamics under which a random amount of mtDNA is lost at each division. This picture could model, for example, mtDNA dynamics in budding yeast, where cells with $50-200$ mtDNA molecules \cite{solieri2010mitochondrial} undergo asymmetric partitioning, with $10-20\%$ of their mitochondrial content being lost at budding events, and homeostasis acting to maintain copy number \cite{rafelski2012mitochondrial}. Fig. \ref{relaxed} illustrates the behaviour of this system with $m_{opt} = 100, \eta = 15$. Interestingly in this case, the variance of copy number reaches a minimum at an intermediate point in each cell cycle, after the partitioning event (which increases the variance) and before an extended period of dynamics under homeostasis has led to an increase in copy number and variance. As above, the difference between $m_{opt}$ and $\mathbb{E}(m, \tau)$ can easily be computed in the long time limit (see Appendix):

\begin{equation}
\mathbb{E}(m, \tau) = m_{opt} - \frac{\eta}{e^{\beta \tau} - 1}.
\end{equation}

\iainnote{The expression for the variance in this case involves derivatives of the $q$-Pochhammer symbol, which at first would seem to prevent extracting simple and intuitive expressions for the variance. However, as with several applications of this approach, the derivatives involved all reduce to simple algebraic expressions (see Appendix).} Taking the limit of many cell divisions $n \rightarrow \infty$, we obtain

\begin{equation}
\mathbb{V}(m,t) = m_{opt} + \frac{\eta}{2( e^{2 \beta \tau} - 1)} \left( 3 e^{2 \beta(\tau - t)} - 2 e^{\beta (\tau - t)} - 2 e^{\beta (2 \tau - t)} \right).
\end{equation}

This expression allows us to characterise the form of the variance curve. Writing $\eta' \equiv \eta/(2(e^{2 \beta \tau} - 1))$, we find that $\mathbb{V}(m, 0) = m_{opt} + \eta' (e^{2 \beta \tau} - 2e^{\beta \tau})$, $\mathbb{V}(m, \tau) = m_{opt} + \eta' (1 - 2 e^{\beta \tau})$, and that the minimum variance occurs at $t' = \tau - 1/\beta \ln \left( \frac{1}{3} (1 + e^{\beta \tau}) \right)$ and takes value $\mathbb{V}(m, t') = m_{opt} - (\eta / 6) \coth (\beta \tau / 2)$. All these results agree exactly with stochastic simulation, as illustrated in Fig. \ref{relaxed}, and other statistics and quantities of interest can readily be extracted from the generating function.

\iainnote{We note that as this example falls in the regime where $\eta \ll m_{opt}$, so the probability of low copy number is negligible, the statistics derived using our generating function approach are reliable (and the approach can also be used to compute the probability distribution function and other moments). In cases where low copy numbers are likely, caution must be taken in employing this approach, as described above.}


\subsection*{Extinction probabilities under balanced copy number dynamics}

\iainnote{A strength of the use of generating functions to analyse stochastic dynamics and partitioning of cellular species is that statistics other than low-order moments can be straightforwardly computed. As demonstrated in the previous subsection, full probability distribution functions can be extracted for cellular populations of agents from generating functions, although these functions can be rather complicated. As a simpler example of biological interest, we here consider the extinction probability $P(0,t)$ of a cellular species, under the two dynamic regimes we have previously considered in the context of mtDNA dynamics. The first example is birth-death dynamics with binomial partitioning, with $\lambda = \ln 2 / \tau + \nu$, as used in the bottleneck section. The second is immigration-death dynamics with binomial partitioning, as used in the relaxed replication section. Both of these examples exhibit a balanced mean copy number, with the expected production of agents over a cell cycle balancing the expected loss through cell divisions. As previously described, the variance of the birth-death case increases with time, whereas the variance of the immigration-death case converges.}


The fact that extinction probability can be straightforwardly extracted from our generating functions, using $P(m=0) = G(0,t)$, allows us to explore the probability of extinction under these dynamics. The resulting expressions under birth-death (BD) and immigration-only (I) models are

\begin{eqnarray}
P_{BD}(0, t) & = & \left( \frac{2^{t/\tau}( \nu \tau(n+2) + n \ln 2) - 2 \nu \tau}{2^{t/\tau} (\nu \tau(n+2) + (n+2) \ln 2) - 2\nu \tau} \right)^{m_0}; \\
P_{I}(0, t) & = & (1 - 2^{-n} u^n u')^{m_0} \exp \left( \frac{ m_{opt} (2^{-n} (u-1)u^n - (u-2) u' - 1)}{u' (u-2)} \right),
\end{eqnarray}

where $u \equiv e^{-\beta \tau}$ and $u' \equiv e^{-\beta t}$. Consideration of the $n \rightarrow \infty$ limit shows that in the long time limit, extinction probability under the birth-death model converges to unity, whereas in the immigration-only model a limiting probability is reached. Setting $t=0$ for simplicity (thus considering the population at the start of a cell cycle), this limiting probability is $\exp ( -m_{opt} (1 + (u-2)^{-1}) )$. The difference between the ID and B cases is due to the irreversibility of extinction under the birth-death model (and the non-zero probability associated with extinction during every cell cycle); by contrast, extinction in the immigration model can be escaped as immigration creates more agents in the cell without necessitating a nonzero source population. Hence, a nonzero probability flux away from the $m=0$ state exists, and eventually balances the flux into that state due to partitioning noise: the extinction probability may thus be thought of as representing the proportion of time during which the system occupies the $m=0$ state.

Further results can straightforwardly be extracted from our formalism for extinction probabilities in non-balanced cases. As a brief example, we consider the birth-death model with binomial partitioning, with a new parameter $\kappa = \lambda - \nu - \ln 2 / \tau$, so that $\kappa$ measures the `excess' birth rate beyond that required for copy number balance. For $\kappa \leq 0$, extinction is certain in the long-time limit, but for $\kappa > 0$ there is a finite probability that the copy number will never reach zero. To illustrate the qualitative behaviour of the system, we set $\nu = 0$, and we obtain 

\begin{equation}
P_{BD}(m=0) = \left( \frac{1 - e^{\kappa n \tau} }{1 + e^{\kappa n \tau} - 2 e^{\kappa (n+1) \tau } } \right)^{m_0},
\end{equation}

showing that extinction probability in the long-time limit decays roughly exponentially with $\kappa$. The more general extinction probabilities for $\nu \not= 0$ and in the absence of cell divisions are given in the Appendix.


\begin{figure}
\includegraphics[width=15cm]{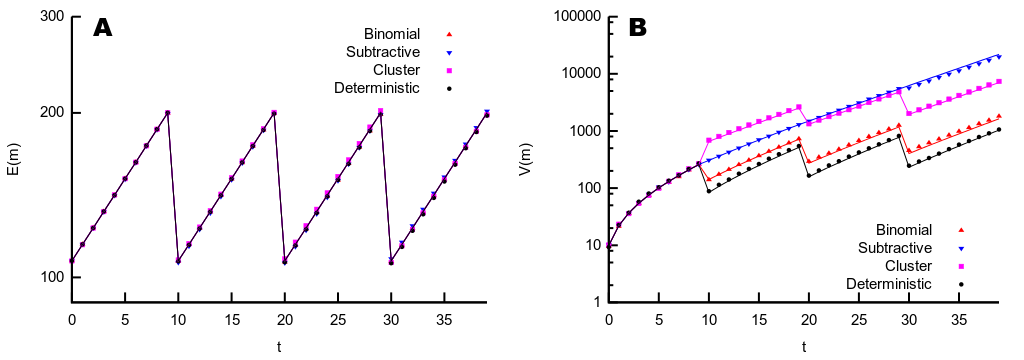}
\caption{\textbf{Solutions from recurrence relations for different partitioning regimes.} Points are from stochastic simulation, lines are theoretical predictions. Deterministic, clustered, subtractive and binomial inheritance regimes are compared. Parameters used were $m_0 = 200, \nu = 0.01, \tau = 10, n_c = 10, \lambda = \log 2 / \tau + \nu, \eta = m_0$ -- the choice of the latter two parameters was fixed to preserve a constant mean copy number. Stochastic simulation results are from an ensemble of $10^5$ simulations of each situation.}
\label{recurillus}
\end{figure}

\subsection*{Other inheritance dynamics}

In addition to the binomial partitioning and random additive or subtractive changes of copy number, we have explored several other possible dynamic regimes of inheritance. The case where a fixed, deterministic number of agents is gained or lost at cell divisions is analytically tractable (see Appendix). We also consider deterministic halving of copy number, so that each daughter cell inherits half of the mother cell's content (rounded down). Additionally, we consider the inheritance of clusters of agents, such that agents are split into clusters of size $n_c$ and these clusters are binomially partitioned. We have not found closed-form analytic solutions for the generating function for a general number of cell divisions for these latter two cases, but analytic statistics can nonetheless be obtained for a given number of cell divisions through the calculation of the appropriate recurrence relations. 

Fig. \ref{recurillus} illustrates the use of this approach to calculate the mean and variance of copy number for these systems, and for birth-death dynamics in the binomial and constant subtractive inheritance regimes described earlier. The agreement between stochastic simulation and analytic results is again excellent, showing, as expected, that deterministic inheritance leads to the lowest magnitude of stochasticity in copy number, followed by binomial partitioning, followed by clustered partitioning (illustrated for $n_c = 10$ in this case). We expect that other inheritance regimes of interest may be addressable through a similar approach.
 
\section{Discussion}
We have introduced a general mathematical formalism with which to address the stochastic dynamics of cellular agents that are inherited according to non-trivial and potentially stochastic dynamics at cell divisions. Our approach differs from, and extends, several previous tools designed to address stochastic partitioning in biology. First, our approach yields full, closed-form generating functions for several cases, allowing the extraction of all details of copy number distributions, rather than focussing on variance or other low-order moments alone. Second, we nowhere assume that a steady-state or equilibrium has been reached, and are thus capable of extracting copy number statistics at any given time during the stochastic biological process of interest. Third, we focus on birth-immigration-death dynamics rather than the more common stochastic gene expression dynamics, with a view to modelling the behaviour of non-protein cellular components (including mtDNA, which we explore in particular). Fourth, we explore several specific partitioning regimes, obtaining closed-form results for arbitrary numbers of cell divisions under those we term binomial partitioning and random and deterministic subtractive or additive inheritance. We also obtain results for finite numbers of cell divisions under deterministic partitioning and binomial partitioning of clusters.

We have focussed on agents undergoing birth-immigration-death (BID) dynamics between cell divisions; we expect that any random dynamics with a corresponding generating function of the form Eqn. \ref{gfsolutionmaintext} will also admit treatment using this approach, paving the way for further generalisation of this approach.

\iainnote{We note that the results arising from our analysis should be interpreted as ensemble statistics of single-cell measurements. If tracking a particular lineage of dividing cells, it should be remembered that the statistics of daughter cells will exhibit correlations due to their shared heritage. Our results represent expected statistics from a well-mixed bulk case.}

In the absence of immigration dynamics, we have shown that a system governed by different rate parameters at different times can also be solved analytically. The rates associated with the production and destruction of cellular agents can vary arbitrarily as long as the rate of this change is lower than the cell division rate. This approach therefore provides a way to explore the statistics of stochastic systems with arbitrarily-changing population size, by contrast with many results from classical statistical genetics which assume a constant or constant-mean population size (and additionally often only provide equilbrium results, preventing the quantitative exploration of systems before a steady state is reached) \cite{wright1942statistical, kimura1955solution, maruyama1980genetic, iizuka2001effective}. It is also straightforward to vary the cell cycle length $\tau$ in these dynamic phases and so allow a treatment of cellular dynamics under varying division times. We have illustrated a general use of this approach in addressing the mtDNA bottleneck, and it has been used to explore the bottleneck specifically in mice in detail \cite{usrecent}. We believe that this formalism may prove useful in other contexts where organellar content is subject to dramatic and non-random population size changes, for example, in considering cellular populations during tumour development, where variability in cellular conditions causes time differences in physiological rate constants as tumour cells continually divide \cite{nicolson1987tumor}.

\iainnote{We have demonstrated the applicability of our stochastic formalism with some illustrative problems from cellular biology. We have found expressions for the cell-to-cell variability in mtDNA populations due to the imposition of a copy number bottleneck of given size, and extended a the classic `relaxed replication' model of mtDNA to include the stochastic dynamics of individual mtDNAs, and the effects of cell divisions. This model is widely influential in the study of mtDNA genetics and disease, but its quantitative analysis has typically been limited to descriptions of its mean behaviour or simulation studies focussed on used. We have thus used our approach to further analytic understanding of this important model.  Furthermore, we have explored in detail the statistics of populations of cellular agents under passive copy number balance over many cell divisions, a situation of importance for organelles and which may be of general applicability in cell biology. }

Accurate models for the variability of cellular populations enable more powerful inference using experimental measurements of mean and variance across cells \cite{johnston2014efficient}. We hope that our results for stochastic inheritance dynamics will facilitate the strengthening of this link between theoretical and experimental biology and allow more information about underlying cellular dynamics to be obtained from the wealth of experimental measurements currently appearing.



\bibliographystyle{unsrt}
\bibliography{partitionnew}

\clearpage
\onecolumngrid
\appendix

\section*{\Large Appendix}

In this Appendix we present some of the lengthier mathematical results used in the main text. We include a Mathematica notebook as part of this article, containing the following derivations.

\section*{Generating function for the BID process}
The derivation of the generating function for BID dynamics in the absence of cell divisions is well known: we include it here for completeness. We write the PDE describing the generating function $G(z,t)$ in Laplace form:

\begin{eqnarray}
\frac{\partial G(z, t)}{dt} - ( \nu (1-z) + \lambda(z^2 - z) ) \frac{\partial G(z, t)}{\partial z} & = &  \alpha (z-1) G  \\
G(z,0) & = & z^{m_0}, \\
\end{eqnarray}

We proceed by using the method of characteristics, writing down ODEs describing how the parameters of $G$, and $G$ itself, changes along a characteristic curve, with progress along such a curve parameterised by $s$. The corresponding ODEs are

\begin{eqnarray}
\frac{dt}{ds} & = & 1 \label{chart} \\
\frac{dz}{ds} & = & - (\nu (1-z) + \lambda (z^2 - z)) \label{charz} \\
\frac{dG}{ds} & = & \alpha (z-1) G \label{charg}
\end{eqnarray}

Eqn. \ref{chart} lets us immediately set $t = s$, omitting a constant of integration as the absolute value of progress along a characteristic curve is unimportant. Using $t = s$ throughout, Eqn. \ref{charz} is solved by

\begin{equation}
z = \frac{1 - \nu e^{c_1 (\lambda - \nu) - t(\lambda - \nu)}}{1 - \lambda e^{c_1 (\lambda - \nu) - t (\lambda - \nu)}} \label{zexp}
\end{equation}

where $c_1$ is a constant of integration, the explicit form of which will be useful later. Rearranging this into an expression for $c_1$ gives

\begin{equation}
c_1 = t + \frac{\ln \left( \frac{z-1}{\lambda z - \nu} \right) }{\lambda - \nu}. \label{c1exp}
\end{equation}

Finally, Eqn. \ref{charg} with Eqn. \ref{zexp} gives us

\begin{equation}
G = c_2 e^{-\alpha t} \left( e^{\lambda t + c_1 \nu} - \lambda e^{\lambda c_1 + \nu t} \right)^{\alpha/\lambda}. \label{intermediategsolnsi}
\end{equation}

$c_2$ is a function of $c_1$ because the quantity $c_1$, the constant of integration acquired when integrating $z$ with respect to $s$, is independent of $s$, and hence forms an independent parameter when integrating $G$ with respect to $s$. We require that $G(t=0) = z^{m_0}$, so we choose

\begin{equation}
c_2(c_1) = \left( e^{c_1 \nu} - \lambda e^{c_1 \lambda} \right)^{-\alpha / \lambda} \left( \frac{ \nu e^{(\lambda - \nu) c_1} - 1}{\lambda e^{(\lambda - \nu) c_1} - 1} \right)^{m_0},
\end{equation}

where the first term cancels the final term in Eqn. \ref{intermediategsolnsi} when $t=0$, and the final term can be seen to extract a factor $z^{m_0}$ from Eqn. \ref{c1exp} for $c_1$ when $t = 0$. We then have

\begin{equation}
G(z,t) = c_2(c_1(z,t)) e^{-\alpha t} \left( e^{\lambda t + c_1(z,t) \nu} - \lambda e^{\lambda c_1(z,t) + \nu t} \right)^{\alpha/\lambda}
\end{equation}

which, after inserting Eqn. \ref{c1exp} and some algebra, gives

\begin{eqnarray}
G(z,t) & = & \left( \frac{ \nu - \lambda }{\lambda e^{(\lambda-\nu)t} (z-1) - \lambda z + \nu} \right)^{\frac{\alpha}{\lambda}} \left( \frac{\nu e^{(\lambda-\nu)t} (z-1) - \lambda z + \nu}{\lambda e^{(\lambda-\nu)t} (z-1) - \lambda z + \nu} \right)^{m_0}  \label{nodivsolnsi} \\
& \equiv & \xi(z,t) (g(z,t))^{m_0}
\end{eqnarray}

\section*{Recurrence relations arising from induction over cell divisions}
This section focusses on the solution of recurrence relations of the form 

\begin{equation}
\zeta_i = \frac{a \zeta_{i+1} + b}{c \zeta_{i+1} + d},\,\text{or equivalently, }\, \zeta_{i+1} = \frac{d \zeta_i - b}{- c \zeta_i + a} \label{recurintro}
\end{equation}

In the Main Text, both $h_i$ and $z_i$ follow relationships of this kind; we use the symbol $\zeta_i$ here to emphasise that the same solution strategy applies in both cases, and describe specific solutions below. This system is solved, after \cite{brand1955sequence}, by defining $\alpha  \equiv \frac{a+d}{c}$, $\beta \equiv \frac{D}{c^2}$, $D \equiv ad - bc$, $y_i = \zeta_i + \frac{d}{c}$ and implicitly defining $w_i$ through $y_i = \frac{w_{i}}{w_{i+1}}$. These changes of variables allow us to find an expression for $w_i$, which can then be substituted back through the above chain to find $\zeta_i$. We have

\begin{eqnarray}
y_i &=  & \alpha - \frac{\beta}{y_{i+1}} \\
\frac{w_i}{w_{i+1}} & = & \alpha - \frac{\beta w_{i+2}}{w_{i+1}} \\
&\rightarrow &\beta w_{i+2} - \alpha w_{i+1} + w_i = 0,
\end{eqnarray}

which is solved by considering solutions to the characteristic equation $\beta k^2 - \alpha k + 1 = 0$, which are straightforwardly $k_{1,2} = \frac{1}{2 \beta} ( \alpha \pm \sqrt{\alpha^2 - 4 \beta} )$. Then

\begin{eqnarray}
w_i & = & C_1 k_1^i + C_2 k_2^i \\
y_i & = & \frac{C_0 k_1^i + k_2^i}{C_0 k_1^{i+1} + k_2^{i+1}} \\
\zeta_i & = & \frac{C_0 k_1^i + k_2^i}{C_0 k_1^{i+1} + k_2^{i+1}} - \frac{d}{c}  \label{solvedrecur}
\end{eqnarray}

where $C_i$ are constants to be determined from boundary conditions. If the boundary condition takes the form $\zeta_n = \frac{p z + q}{r z + s}$, as is the case throughout the situations we consider, we obtain

\begin{eqnarray}
\frac{C_0 k_1^i + k_2^i}{C_0 k_1^{i+1} + k_2^{i+1}} - \frac{d}{c} & = & \frac{p z + q}{r z + s} \\
\Rightarrow C_0 & = & \frac{k_2^n k_1^{-n} (k_2 c (p z + q) + k_2 d (r z + s) - c(r z + s)) }{c (rz + s) - k_1 c (pz + q) - k_1 d (rz + s)}. \label{c0eqnsi}
\end{eqnarray}

Thus, given knowledge of $a, b, c, d$ from the recurrence relation and $p, q, r, s$ from the initial condition, we can obtain $k_1, k_2$ through $\alpha$ and $\beta$ and hence use Eqns. \ref{c0eqnsi} and \ref{solvedrecur} to obtain a solution to the recurrence relation. Below, we use this approach to obtain solutions for the systems of interest in the main text.

\subsection*{Binomial partitioning solution for $h$}
We will use the substitutions $l = e^{(\lambda - \nu) \tau}, l' = e^{(\lambda-\nu)t}$. The original recurrence relation is

\begin{eqnarray}
h_i & = & g \left( \frac{1}{2} + \frac{h_{i+1}}{2}, \tau \right) \label{recur1} \\
& = & \frac{(\nu l - \lambda) h_{i+1} + ( - \lambda - \nu( l - 2))} { (\lambda (l-1)) h_{i+1} + (2 \nu - \lambda (l+1))} \\
h_n & = & g (z, t) \\
& = & \frac{ (\nu l' - \lambda) z + (\nu - \nu l')}{(\lambda(l'-1)) z + (\nu - \lambda l')};
\end{eqnarray}

hence, $a = (\nu l - \lambda), b = (2 \nu - \nu l - \lambda), c = (\lambda l - \lambda), d = (2 \nu - \lambda - \lambda l), p = (\nu l' - \lambda), q = (\nu - \nu l'), r = (\lambda(l'-1)), s = (\nu - \lambda l')$. Using these values to determine $\alpha, k_1, k_2$, and after some algebra, we obtain

\begin{equation}
h_i = \frac{2^i l^n l' (z-1) (\lambda + \nu (l-2)) + 2^n l^i ( \lambda (l' - z(l + l' - 2)) + \nu (l-2) )}{2^i l^n l' \lambda (l-1) (z-1) + 2^n l^i ( \lambda (l' - z(l + l' - 2)) + \nu (l-2) )}. \label{hiappendixbinom}
\end{equation}

\subsection*{Subtractive partitioning solution for $h$}
\begin{eqnarray}
h_i & = & g(h_{i+1}, \tau) \\
& = & \frac{ (\nu l - \lambda) h_{i+1} + (\nu - \nu l)}{(\lambda(l-1)) h_{i+1} + (\nu - \lambda l)}; \\
h_n & = & g(z, t), \label{telorecur2} \\
& = & \frac{ (\nu l' - \lambda) z + (\nu - \nu l')}{(\lambda(l'-1)) z + (\nu - \lambda l')};
\end{eqnarray}

hence $a = (\nu l - \lambda), b = (\nu - \nu l), c = (\lambda (l-1)), d = (\nu - \lambda l), p = (\nu l' - \lambda), q = (\nu - \nu l'), r = \lambda (l'-1), s = (\nu - \lambda l')$. Then

\begin{equation}
h_i = \frac{l^n l' \nu (z-1) + l^i (\nu - \lambda z)} {l^n l' \lambda (z-1) + l^i (\nu - \lambda z)} \label{subtractivehsolnsi}
\end{equation}

\subsection*{Binomial partitioning solution for $z$}

\begin{eqnarray}
z_i & = & \frac{1}{2} + \frac{g(z_{i-1}, \tau)}{2} \\
& = & \frac{(l(\lambda + \nu) - 2 \lambda) z_{i-1} + (2 \nu - l (\lambda + \nu)) }{ (2 \lambda (l-1)) z_{i-1} + (2 \nu - 2 \lambda l)} \\
\Rightarrow z_i & = & \frac{ (2 \nu - 2 \lambda l) z_{i+1} + (l (\lambda + \nu) - 2 \nu) } { (2 \lambda (1 - l)) z_{i+1} + (l (\lambda + \nu))} \label{switchrecur} \\
z_1 & = & \frac{1}{2} + \frac{g(z,t)}{2}. \\
& = & \frac{(l'(\lambda + \nu) - 2 \lambda) z + (2 \nu - l' (\lambda + \nu)) }{ (2 \lambda (l'-1)) z + (2 \nu - 2 \lambda l')}.
\end{eqnarray}

In Eqn. \ref{switchrecur} we have used the equivalence in Eqn. \ref{recurintro} to rewrite the recurrence relation in the form we have previously solved. Subsequently, $a = (2 \nu - 2 \lambda l) , b = (l (\lambda + \nu) - 2 \nu) , c = (2 \lambda (1 - l)), d = (l (\lambda + \nu)), p = l'(\lambda + \nu) - 2 \lambda) , q = (2 \nu - l' (\lambda + \nu)), r = (2 \lambda (l'-1)), s = (2 \nu - 2 \lambda l')$, leading to

\begin{equation}
z_i = \frac{l^i l' (z-1) (l (\lambda + \nu) - 2 \nu) - 2^i l (\lambda l' (z-1) + (l-2) (\lambda z - \nu))}{2 \lambda l^i l' (z-1)(l-1) - 2^i l (\lambda l' (z-1) + (l-2) (\lambda z - \nu))} \label{ziappendixbinom}
\end{equation}

\subsection*{Subtractive partitioning solution for $z$}

\begin{eqnarray}
z_i & = & g(z_{i-1},\tau), \\
& = & \frac{ (\nu l - \lambda) z_{i-1} + (\nu - \nu l)}{(\lambda(l-1)) z_{i-1} + (\nu - \lambda l)}; \\
\Rightarrow z_i & = & \frac{ (\nu - \lambda l) z_{i+1} + (\nu l - \nu) } { (\lambda (1-l)) z + (\nu l - \lambda)} \\
z_1 & = & g(z,t), \\
& = & \frac{ (\nu l' - \lambda) z + (\nu - \nu l')}{(\lambda(l'-1)) z + (\nu - \lambda l')};
\end{eqnarray}

Hence $a = (\nu - \lambda l), b = (\nu l - \nu), c = (\lambda (1-l)), d = (\nu l - \lambda), p = l'(\lambda + \nu) - 2 \lambda) , q = (2 \nu - l' (\lambda + \nu)), r = (2 \lambda (l'-1)), s = (2 \nu - 2 \lambda l')$, leading to

\begin{equation}
z_i = \frac{l^i l' \nu (z-1) + l (\nu - \lambda z)}{l^i l' \lambda (z-1) + l (\nu - \lambda z)} \label{ziappendixsub}
\end{equation}

\section*{Products of prefactors over the inductive process}

We are concerned with an expression for the product in Eqn. \ref{generalgf}, $\prod_{i=1}^n \phi (z_i, \tau)$. We first consider $\prod_{i=1}^n \xi (z_i,\tau)$ which occurs as a factor in this expression in every partitioning regime. We recall that $\xi(z, t)$ has the form

\begin{equation}
\xi(z, t) = \left( \frac{ \nu - \lambda} {\lambda e^{(\lambda - \nu) t} (z-1) - \lambda z + \nu} \right)^{\alpha / \lambda}
\end{equation}

It will be convenient to write 

\begin{eqnarray}
z_i & = & \frac{\tilde{A}_1 \rho_A^i + \tilde{B}_1 \rho_B^i}{\tilde{A}_2 \rho_A^i + \tilde{B}_2 \rho_B^i} \\
\xi(z_i, \tau) & = & \left( \frac{A_1 \rho_A^i + B_1 \rho_B^i}{A_2 \rho_A^i + B_2 \rho_B^i} \right)^{\gamma}, \label{rationalform}
\end{eqnarray}

where $A_1 = (\tilde{A}_2(\nu - \lambda)), B_1 = (\tilde{B}_2 (\nu - \lambda)), A_2 = ( \lambda l (\tilde{A}_1 - \tilde{A}_2) + \nu \tilde{A}_2 - \lambda \tilde{A}_1), B_2 = ( \lambda l (\tilde{B}_1 - \tilde{B}_2) + \nu \tilde{B}_2 - \lambda \tilde{B}_1)$, $\gamma = \alpha / \lambda$, following from the form of $\xi(z, t)$.

The product of an expression of this form can be written

\begin{equation}
\prod_{i=1}^n \xi(z_i,t_i) \equiv \prod_{i=1}^{n} \left( \frac{A_1 \rho_A^i + B_1 \rho_B^i}{A_2 \rho_A^i + B_2 \rho_B^i} \right)^{\gamma} = \left( B_1^n \rho_B^{in} \prod_{i=1}^n \left( 1 + \frac{A_1}{B_1} \frac{\rho_A^i}{\rho_B^i} \right) \right)^{\gamma} \left( B_2^n \rho_B^{in} \prod_{i=1}^n \left( 1 + \frac{A_2}{B_2} \frac{\rho_A^i}{\rho_B^i} \right) \right)^{-\gamma}
\end{equation}

Here we make use of the $q$-Pochhammer symbol $(a;q)_n$, defined by

\begin{equation}
(a;q)_n \equiv \prod_{k=0}^{n-1} (1 - aq^k).
\end{equation} 

It can then be seen, by setting $a = -A_j / B_j$ and $q = \rho_A / \rho_B$, 

\begin{eqnarray}
B^n \rho_B^{in} \prod_{i=1}^n \left( 1 + \frac{A_j}{B_j} \frac{\rho_A^i}{\rho_B^i} \right) & \equiv & B^n \rho_B^{in} \frac{1}{1 + A_j/B_j} (-A_j/B_j; \rho_A/\rho_B)_{n+1}, \\
& = & B_i^{n+1} \rho_B^{in} \frac{(-A_j/B_j; \rho_A/\rho_B)_{n+1}}{A_j + B_j}
\end{eqnarray}

and hence that

\begin{equation}
\left( B_1^n \rho_B^{in} \prod_{i=1}^n \left( 1 + \frac{A_1}{B_1} \frac{\rho_A}{\rho_B} \right) \right)^{\gamma} \left( B_2^n \rho_B^{in} \prod_{i=1}^n \left( 1 + \frac{A_2}{B_2} \frac{\rho_A}{\rho_B} \right) \right)^{-\gamma} \equiv \left( \frac{B_1^{n+1} B_2^{-n-1} (A_2 + B_2) (-A_1/B_1; \rho_A/\rho_B)_{n+1}} {(A_1+B_1) (-A_2/B_2; \rho_A/\rho_B)_{n+1}} \right)^{\gamma}, \label{siproductform1}
\end{equation}

yielding a simple form for the product of interest. For the binomial case, we have from Eqn. \ref{ziappendixbinom}:

\begin{equation}
z_i = \frac{l^i l' (z-1) (l (\lambda + \nu) - 2 \nu) - 2^i l (\lambda l' (z-1) + (l-2) (\lambda z - \nu))}{2 \lambda l^i l' (z-1)(l-1) - 2^i l (\lambda l' (z-1) + (l-2) (\lambda z - \nu))} \label{zilinkbinom}
\end{equation}

so that $\tilde{A}_1 = (l' (z-1) (l (\lambda + \nu) - 2 \nu)), \tilde{B}_1 = (-l (\lambda l' (z-1) + (l-2) (\lambda z - \nu))), \tilde{A}_2 = (2 \lambda l' (z-1) (l-1)), \tilde{B}_2 = (-l (\lambda l' (z-1) + (l-2) (\lambda z - \nu)))$, hence $A_1 = 2 \lambda l' (l-1) (z-1) (\nu - \lambda), A_2 = \lambda l l' (l-1) (z-1) (\nu - \lambda), B_1 = B_2 = -l ( \lambda l' (z-1) + (l-2) (\lambda z - \nu) )(\nu - \lambda)$, and $\rho_A = l, \rho_B = 2, \gamma = \alpha/\lambda$.

For the subtractive case, we have from Eqn. \ref{ziappendixsub}:

\begin{equation}
z_i = \frac{l^i l' \nu (z-1) + l (\nu - \lambda z)}{l^i l' \lambda (z-1) + l (\nu - \lambda z)} \label{zilinksub}
\end{equation}

so that $\tilde{A}_1 = (l' \nu (z-1)), \tilde{B}_1 = (l (\nu - \lambda z)), \tilde{A}_2 = (l' \lambda (z-1)), \tilde{B}_2 = (l (\nu - \lambda z))$, hence $A_1 = \lambda l'  (z-1) (\nu - \lambda), A_2 = \lambda l l'  (z-1) (\nu - \lambda), B_1 = B_2 = l (\nu - \lambda) (\nu - \lambda z)$, and $\rho_A = l, \rho_B = 1, \gamma = \alpha/\lambda$.

\subsection*{Other products}

We can also use this result to compute the product of exponentiated prefactors involved in the subtractive inheritance regimes. The first product required is $\prod_{i=1}^n g(z_i, \tau)^{-\eta}$. We recall the definitions of the recurrence relations in the Main Text

\begin{eqnarray}
h_i(z,t) & = & g \left( \theta( h_{i+1}) , \tau  \right) \,;\, h_n(z,t) = g (z, t) \\
z_{i+1} & = & \theta(g(z_i, \tau)) \,;\, z_1 = \theta(g(z, t)). 
\end{eqnarray}

For both subtractive inheritance cases, $\theta(g(z,t)) = g(z,t)$, so it can straightforwardly be seen that

\begin{eqnarray}
z_{i+1} & = & g(z_i, \tau) \,;\, z_1 = g(z,t) \\
h_i(z,t) & = & g( h_{i+1}, \tau ) \,;\, h_n(z,t) = g(z,t) \\
\text{and so}\,\, h_n(z,t) & = & z_1 \,;\, h_{n-i+1} = z_i. \label{equivrecur}
\end{eqnarray}

Thus, the product $\prod_{i=1}^n g(z_i, \tau)^{-\eta}$ is equivalent to the product $\prod_{j=1}^n h_j^{-\eta}$, where $j = n - i + 1$. As $i$ and $j$ are dummy variables, we can then identify the required solution as $\prod_{i=1}^n h_i^{-\eta}$. We have from Eqn. \ref{subtractivehsolnsi} that

\begin{eqnarray}
h_i & = & \frac{l^n l' \nu (z-1) + l^i (\nu - \lambda z)} {l^n l' \lambda (z-1) + l^i (\nu - \lambda z)} \\
& = & \frac{\nu l' x_1^i (-x_2)^n (z-1) + x_1^n (-x_2)^i (\nu - \lambda z)} {\lambda l' x_1^i (-x_2)^n (z-1) + x_1^n (-x_2)^i (\nu - \lambda z)} \label{thirdproduct1}
\end{eqnarray}

where we have rewritten the final line to avoid a diverging factor of $(z-1)^{-1}$ appearing in the Pochhammer symbol, using

\begin{eqnarray}
x_1 & = & \frac{\lambda}{\lambda - \nu} (e^{(\nu - \lambda) \tau} - 1) \\
x_2 & = & \frac{\lambda}{\lambda - \nu} (e^{(\lambda - \nu) \tau} - 1) 
\end{eqnarray}

We then see that $h_i^{-\eta}$ is of the form Eqn. \ref{rationalform}, so we can use the result therein, with $A_1 = \nu l' (-x_2)^n (z-1), B_1 = B_2 = x_1^n (\nu - \lambda z), A_2 = \lambda l' (-x_2)^n (z-1), \rho_A = x_1, \rho_B = (-x_2), \gamma = -\eta$.

Finally, we wish to compute an expression for $\prod_{i=1}^n \left( \frac{1}{2} + \frac{1}{2 g(z_i,t)} \right)^{2 \eta} $, of use in the random subtractive regime. We again exploit the relation between $g(z_i, \tau)$ and $h_j$ in Eqn. \ref{equivrecur} to show that the kernel of the desired product is equivalent to $\left( \frac{1}{2} + \frac{1}{2 h_i} \right)$. Again, we use $h_i$ from Eqn. \ref{subtractivehsolnsi}; after some algebra this expression reduces to

\begin{equation}
\frac{l' x_1^i (-x_2)^n (z-1) (\lambda - \nu) + 2 x_1^n (-x_2)^i (\nu - \lambda z)}{2 l' x_1^i (-x_2)^n \nu (z-1) + 2 x_1^n (-x_2)^i (\nu - \lambda z)}, \label{thirdproduct2}
\end{equation}

whereupon we can use Eqn. \ref{siproductform1} with $A_1 = l' (-x_2)^n (z-1) (\lambda - \nu), B_1 = B_2 = 2 x_1^n (\nu - \lambda z), A_2 = 2 l' (-x_2)^n \nu (z-1), \rho_A = x_1, \rho_B = (-x_2), \gamma = 2 \eta$.

\section*{Full forms of generating functions}


To recap, we use $\alpha, \lambda, \nu$ to respectively represent the rates of immigration, birth, and death in our model; $m_0$ for initial copy number; $\tau$ for cell cycle length, $n$ for the number of divisions that have occurred, and $t$ for the elapsed time since the most recent cell division. We employ simplifying symbols $l \equiv e^{(\lambda - \nu) \tau}; l' \equiv e^{(\lambda - \nu)t}$ and $x_1 \equiv \lambda (l^{-1} - 1) / (\lambda - \nu); x_2 \equiv \lambda (l-1) / (\lambda - \nu)$.

The general form of the generating functions was shown in the Main Text to be

\begin{equation}
G_n(z,t) = \underbrace{ \left( \prod_{i=1}^n \xi(z_i, \tau) \right)}_{\text{(i)}} \underbrace{ \left( \prod_{i=1}^n \phi(z_i, \tau) \right) }_{\text{(ii)}} \underbrace{ \vphantom{\prod_{i=1}^n} \xi(z,t) }_{\text{(iii)}} \underbrace{ \vphantom{\prod_{i=1}^n} h_0(z,t)^{m_0} }_{\text{(iv)}},
\end{equation}

where $z_i$ and $h_i$ are the solutions to recursion relations defined in the Main Text. The term (iii) is the same in all calculations and is, from Eqn. \ref{nodivsolnsi},

\begin{equation}
\xi(z,t) = \left( \frac{\nu - \lambda}{\lambda l (z-1) - \lambda z + \nu } \right)^{\alpha / \lambda}. \label{xisi}
\end{equation}

The generating function in the case of binomial partitioning at cell divisions involves (i) Eqn. \ref{siproductform1} applied to the appropriate terms described in Eqn. \ref{zilinkbinom}; (ii) unity; (iii) Eqn. \ref{xisi}; and (iv) $h_0$ from Eqn. \ref{hiappendixbinom}, giving overall


\begin{eqnarray}
G_n(z,t) & = & 
\left( \frac{ (\lambda l l' (l-1)(z-1) - l (\lambda l' (z-1) + (l-2)(\lambda z - \nu) ) ) \left( \frac{-2 \lambda l'(l-1)(z-1)}{-l(\lambda l'(z-1) + (l-2)(\lambda z - \nu))}; \frac{l}{2} \right)_{n+1} }{ ( 2 \lambda l' (l-1) (z-1) - l (\lambda l' (z-1) + (l-2) (\lambda z - \nu))) \left( \frac{-\lambda l l' (l-1) (z-1) }{-l(\lambda l' (z-1) + (l-2)(\lambda z - \nu))}; \frac{l}{2} \right)_{n+1} } \right)^{\alpha / \lambda} \nonumber \\ 
&& \times \left( \frac{\nu - \lambda}{\lambda l (z-1) - \lambda z + \nu } \right)^{\alpha / \lambda} \nonumber \\
&& \times \left( \frac{l^n l'(z-1)(\lambda + \nu(l - 2)) + 2^n (\lambda (l' - z(l + l' - 2)) + \nu (l-2))}{l^n l' (z-1) \lambda (l-1) + 2^n (\lambda(l' - z(l+ l' - 2)) + \nu(l-2))} \right)^{m_0} \label{biggiebinom}
\end{eqnarray}



The generating function in the case of random subtractive inheritance involves (i) Eqn. \ref{siproductform1} applied to the appropriate terms described in Eqn. \ref{zilinksub}; (ii) Eqn. \ref{siproductform1} applied to Eqn. \ref{thirdproduct2} as described; (iii) Eqn. \ref{xisi}; and (iv) $h_0$ from Eqn. \ref{subtractivehsolnsi}, and is

\begin{eqnarray}
G_n(z,t) & = & \left( \frac{ ( 2 l' \nu (-x_2)^n (z-1) - 2 x_1^n (\lambda z - \nu) ) \left( \frac{ -l' (\lambda + \nu) (-x_2)^n (z-1) }{-2 x_1^n (\lambda z - \nu) } ; \frac{x_1}{-x_2} \right)_{n+1} }{ (l' (\lambda + \nu) (-x_2)^n (z-1) - 2 x_1^n (\lambda z - \nu) ) \left( \frac{ 2 l' \nu (-x_2)^n (z-1) }{-2 x_1^n (\lambda z - \nu) } ; \frac{x_1}{-x_2} \right)_{n+1} } \right)^{2 \eta} \nonumber \\
&& \times  \left( \frac{ (\lambda l' l(z-1) + l (\nu - \lambda z)) \left( \frac{-\lambda l'(z-1)}{l(\nu - \lambda z)} ; l \right)_{n+1} }{ (\lambda l'(z-1) + l (\nu - \lambda z)) \left( \frac{ -\lambda l' l (z-1) }{l(\nu - \lambda z)} ; l \right)_{n+1} } \right)^{\alpha / \lambda} \nonumber \\
&& \times \left( \frac{\nu - \lambda}{\lambda l (z-1) - \lambda z + \nu } \right)^{\alpha / \lambda} \left( \frac{l^n l' \nu (z-1) + \nu - \lambda z}{l^n l' \lambda (z-1) + \nu - \lambda z} \right)^{m_0}
\end{eqnarray}


\section*{Birth-death dynamics for the mtDNA bottleneck}
For birth-death dynamics, $\alpha = 0$, so for binomial partitioning the generating function takes the form of the final term in Eqn. \ref{biggiebinom}.

In the case of balanced copy number, with $\lambda = \ln 2 / \tau + \nu$,

\begin{equation}
g(z,t) = \frac{(2^{t/\tau} - 1)(z-1) \nu \tau - z \ln 2}{2^{t/\tau} (z-1)(\nu \tau + \ln 2) - z \ln 2 - \nu \tau (z-1)}
\end{equation}

The corresponding solutions for $z_i, h_i$ are
\begin{eqnarray}
z_i & = & \frac{2^{t/\tau} (z-1) ((i+1)\nu \tau + i \ln 2) - 2(z-1) \nu \tau - z \ln 4}{2^{t/\tau} (z-1)(i+1)(\nu \tau + \ln 2) - 2(z-1) \nu \tau - z \ln 4} \\
h_i & = & \frac{2^{t/\tau} (z-1) ((i-n-2) \nu \tau + (i-n) \ln 2) + 2 \nu \tau (z-1) + z \ln 4}{2^{t/\tau}(z-1)(i-n-2) (\nu \tau + \ln 2) + 2 \nu \tau (z-1) + z \ln 4}
\end{eqnarray}

so

\begin{equation}
G_n(z,t) = \underbrace{ \vphantom{\left(\frac{a^a}{a^a}\right)^a} 1 }_{(i)} \times \underbrace{ \vphantom{\left(\frac{a^a}{a^a}\right)^a} 1 }_{(ii)} \times \underbrace{ \vphantom{\left(\frac{a^a}{a^a}\right)^a} 1 }_{(iii)} \times \underbrace{ \left( \frac{2\nu \tau (z - 1) - 2^{t/\tau} (z-1)((n+2)\nu \tau + n \ln 2) + z \ln 4} {2\nu \tau(z-1) -2^{t/\tau}(z-1)(n+2)(\nu \tau + \ln 2) + z \ln 4} \right)^{m_0} }_{(iv)}.
\end{equation}

\section*{Relaxed replication of mtDNA dynamics}

The generating function for a single cell cycle (involving immigration and death dynamics) can straightforwardly be found

\begin{equation}
G(z,t) = e^{1 - e^{-\beta t}} m_{opt} (z-1) \left( 1 + e^{-\beta t} (z-1) \right)^{m_0},
\end{equation}

and so

\begin{eqnarray}
\xi(z,t) & \equiv & e^{1 - e^{-\beta t}} m_{opt} (z-1) \\
g(z,t) & \equiv & 1 + e^{-\beta t} (z-1).
\end{eqnarray}

For binomial partitioning, we have $\phi(z, t) = 1$ and $\theta(g(z,t)) = (1/2 + g(z,t)/2)$. The solutions of recurrence relations Eqns. \ref{mainrecur1}-\ref{mainrecur2} are then

\begin{eqnarray}
z_i & = & 1 + 2^{-i} e^{-\beta (t + (i-1) \tau)} (z-1) \\
h_i & = & 1 + 2^{i-n} e^{-\beta (t + (n-i) \tau)} (z-1)
\end{eqnarray}

So the overall generating function is

\begin{eqnarray}
G_n(z,t) & = & \underbrace{ \exp \left( m_{opt} (z-1) \left(1 - e^{-\beta t} + \left( \frac{2^{-n} e^{-\beta t} (e^{\beta \tau} - 1) (2^n - e^{-\beta n \tau})}{2 e^{\beta \tau} - 1} \right) \right) \right) }_{\text{(i)}} \underbrace{ \vphantom{\left(\frac{a^a}{a^a}\right)^a} \times 1 \times }_{\text{(ii)}} \underbrace{ \vphantom{\left(\frac{a^a}{a^a}\right)^a} \exp \left( 1 - e^{-\beta t} m_{opt} (z-1) \right) }_{(iii)} \nonumber \\
&& \times \underbrace{ \left( 1 + 2^{-n} e^{-\beta (t + n \tau)} (z-1) \right)^{m_0} }_{(iv)},
\end{eqnarray}

which can be written as 

\begin{equation}
G(z,t) = e^{a z + b} (c z + d)^{m_0}, \label{abcdrr}
\end{equation}

with 

\begin{eqnarray}
a & = & m_{opt} \left( 1 - e^{-\beta t} + \frac{2^{-n} e^{-\beta t} (e^{\beta \tau} - 1)(2^n - e^{-\beta n \tau} )}{2 e^{\beta \tau} - 1} \right); \label{abcdrr1} \\
b & = & -a; \\
c & = & 2^{-n} e^{-\beta (t + n \tau)}; \\
d & = & 1 - c. \label{abcdrr2}
\end{eqnarray}

In particular, the mean copy number is $\left( a e^{az + b} (cz+d)^{m_0} + e^{az+b} m_0 c(cz+d)^{m_0-1} \right)_{z=1}$, giving

\begin{equation}
\mathbb{E}(m,t) = 2^{-n} e^{-\beta (t + n \tau)} m_0 + m_{opt} \left( 1 - e^{-\beta t} + \frac{2^{-n} e^{-\beta t} (e^{\beta \tau} - 1) (2^n - e^{-\beta n \tau})}{2 e^{\beta \tau} - 1} \right)
\end{equation}

and setting $t = \tau$ (at the end of a cell cycle), we obtain 

\begin{equation}
\mathbb{E}(m, \tau) - m_{opt} = \frac{1 + 2^{-n} ( e^{-\beta n \tau} - e^{-\beta (n+1) \tau} )}{2 e^{\beta \tau} - 1} m_{opt} - 2^{-n} e^{-\beta \tau (n+1)} m_0.
\end{equation}

The $n \rightarrow \infty$ limit of this expression is

\begin{equation}
\mathbb{E}(m, \tau) - m_{opt} \xrightarrow{n \rightarrow \infty} \frac{1}{2 e^{\beta \tau} - 1} m_{opt}.
\end{equation}

In this $n \rightarrow \infty$ limit the generating function reduces to

\begin{equation}
G(z,t) \xrightarrow{n \rightarrow \infty} \exp \left( m_{opt} (z-1) \left( 1 - e^{-\beta t}  + \frac{e^{-\beta t} (e^{-\beta t} -1)}{2 e^{\beta \tau} -1} \right) \right),
\end{equation}

so that 

\begin{equation}
P(m,t) \xrightarrow{n \rightarrow \infty} \frac{1}{m!} \left( \frac{m_{opt} (1 - 2e^{\beta \tau} + e^{-\beta(t - \tau)} )}{2 e^{\beta \tau} - 1} \right)^m \exp \left( \frac{m_{opt} (1 - 2e^{\beta \tau} + e^{-\beta(t - \tau)} )}{2 e^{\beta \tau} - 1} \right).
\end{equation}

Using Eqn. \ref{abcdrr} and Leibniz's rule, the general probability distribution function is given by

\begin{eqnarray}
P(m,t) & = & \frac{1}{m!} \left. \frac{\partial^m G}{\partial z^m}\right|_{z=0} \\
& = & \left. \frac{1}{m!} \sum_{k = 0}^m \binom{m}{k} a^{m-k} e^{a z + b} \frac{m_0!}{(m_0 - k)!} (c z + d)^{m_0 - k} \right|_{z=0}\\
& = & \left. \frac{e^{a z + b} (-c)^m (cz + d)^{m_0 - m}}{m!} U \left(-m, 1-m+m_0, \frac{-a (cz+d)}{c} \right)\right|_{z=0}, \\
& = & (1/m!) (-c)^m d^{m_0 - m} e^b U \left(-m, 1-m+m_0, -ad/c \right)
\end{eqnarray}

where $U(a,b,z)$ is the confluent hypergeometric function.

For subtractive partitioning, we have $\phi(z, t) = (1/2 + 1/(2 g(z,t)))^{2 \eta}$ and $\theta(g(z,t)) = g(z,t)$. The solutions of recurrence relations Eqns. \ref{mainrecur1}-\ref{mainrecur2} are then

\begin{eqnarray}
z_i & = & 1 + e^{-\beta (t + (i-1) \tau)}(z-1) \\
h_i & = & 1 + e^{-\beta (t + (n-i) \tau)}(z-1)
\end{eqnarray}

So the overall generating function is

\begin{eqnarray}
G_n(z,t) & = & \underbrace{ \vphantom{\left(\frac{a^a}{a^a}\right)^a} \exp \left( m_{opt} (z-1) \left( e^{-\beta t} - e^{-\beta (t + n \tau)} \right) \right) }_{(i)} \underbrace{ 4^{\eta} \left( \frac{(z - 1 + e^{\beta (t + n \tau)} ) \left( \frac{-1}{2} e^{-\beta (t + n \tau) (z-1)} ; e^{\beta \tau} \right)_{n+1} }{ (z - 1 + 2e^{\beta (t + n \tau)} ) \left( -e^{-\beta (t + n \tau) (z-1)} ; e^{\beta \tau} \right)_{n+1} } \right)^{2 \eta} }_{(ii)} \nonumber \\
&& \times \underbrace{ \vphantom{\left(\frac{a^a}{a^a}\right)^a} \exp \left(m_{opt} (z-1) (1 - e^{-\beta t}) \right) }_{(iii)} \underbrace{ \vphantom{\left(\frac{a^a}{a^a}\right)^a} \left(1 + e^{-\beta (t + n \tau)} (z-1) \right)^{m_0} }_{(iv)}.
\end{eqnarray}

In particular,

\begin{equation}
\mathbb{E}(m,t) = m_{opt} + e^{-\beta (t + n \tau)} (m_0 - m_{opt} + \eta ( 1 + (0; e^{\beta \tau})'_{n+1} ) ) .
\end{equation}

It follows straightforwardly from the definition of the $q$-Pochhammer symbol that 

\begin{eqnarray}
(0; q)'_{n+1} & \equiv & \frac{q^{n+1} - 1}{q - 1} \\
(0; q)''_{n+1} & \equiv & q \frac{(q^n - 1)(q^{n+1} - 1)}{(q+1)(q-1)^2}
\end{eqnarray}

Using these results and setting $t = \tau$ (at the end of a cell cycle), we obtain after some manipulation

\begin{equation}
\mathbb{E}(m, \tau) - m_{opt} = m_0 e^{-\beta \tau (n+1)} + \frac{\eta (1 - e^{-\beta n \tau} ) + m_{opt} ( e^{- \beta (n+1) \tau} + e^{-\beta n \tau} )}{1 - e^{\beta \tau} }
\end{equation}

and the only term retained in the $n \rightarrow \infty$ limit is

\begin{equation}
\mathbb{E}(m, \tau) - m_{opt} \xrightarrow{n \rightarrow \infty} \frac{\eta}{1-e^{\beta \tau}}.
\end{equation}

\section*{Different dynamic phases}

We are concerned with the extension of the generating function for the birth-death process over $n$ cell divisions with the same rate parameters $\lambda, \nu$ to the case where we have different dynamic phases described by parameters $\{ \lambda_1, \nu_1 \}, \{\lambda_2, \nu_2 \}, ...$. The generating function for the birth-death process, without immigration, is simply $h_0$ from Eqn. \ref{binomialhmaintext}. We will write this expression in the form 

\begin{equation}
G(z, t | m_0) = \left( \frac{Pz + Q}{R z + S} \right)^{m_0},
\end{equation}

with coefficients

\begin{eqnarray}
P &= & 2^n \lambda (l + l' -2) -  l^n l' (\lambda + \nu(l-2)) \label{eqnforA} \\
Q & = & l^n l' (\lambda + \nu(l-2)) - 2^n (\lambda l' + \nu (l-2)) \\
R & = & - \lambda l^n l' (l-1) + 2^n \lambda (l + l' -2) \\
S & = & 2 \lambda l^n l' (l-1) - 2^n l (\lambda l' + \nu(l-2)).
\end{eqnarray}

If we now label these coefficients with an index $r$ denoting the appropriate dynamic phase, so that, for example, $P_r$ is Eqn. \ref{eqnforA} with $\lambda_r, \nu_r, n_r$ replacing $\lambda, \nu, n$, we can write:

\begin{eqnarray}
h_{r_{max}} & = & \frac{P_{r_{max}} z + Q_{r_{max}}}{R_{r_{max}} z + S_{r_{max}}} \\
h_r & = & \frac{P_r h_{r+1} + Q_r}{R_r h_{r+1} + S_r} \\
g_{overall} = h_0 & \equiv & \frac{\tilde{P}_0 z + \tilde{Q}_0}{\tilde{R}_0 z + \tilde{S}_0}, \label{tilderecur}
\end{eqnarray}

where $\tilde{P}_r, ... \tilde{S}_r$ are given by the recurrence equations

\begin{eqnarray}
\tilde{P}_r & = & P_r \tilde{P}_{r+1} + Q_r \tilde{R}_{r+1} \\
\tilde{Q}_r & = & Q_r \tilde{Q}_{r+1} + Q_r \tilde{S}_{r+1} \\
\tilde{R}_r & = & R_r \tilde{P}_{r+1} + S_r \tilde{R}_{r+1} \\
\tilde{S}_r & = & S_r \tilde{Q}_{r+1} + S_r \tilde{S}_{r+1}, 
\end{eqnarray}

with $\tilde{P}_n = P_n, ..., \tilde{S}_n = S_n$. If we write the matrix

\begin{equation}
\mathbf{M}_r = \left( \begin{array}{cc}
P_r & Q_r \\
R_r & S_r \end{array} \right) \\
\end{equation}

the general solutions to these equations can compactly be given by

\begin{equation}
 \left( \begin{array}{cc}
\tilde{P}_0 & \tilde{Q}_0 \\
\tilde{R}_0 & \tilde{S}_0 \end{array} \right)  = \prod_{j = 0}^{r_{max}} \mathbf{M}_j
\end{equation}

\section*{Birth-death-binomial extinction probabilities without balance and/or cell divisions}

The birth-death-binomial generating function is given by setting $\alpha = 0$ in Eqn. \ref{biggiebinom}. We set $\lambda = \kappa + \nu + \ln 2 / \tau$ and only consider times immediately after cell divisions, hence setting $t = 0$ and $l' = 1$, giving

\begin{equation}
G_{BD}(z,0) = \left( \frac{ 2 \nu (l_k - 1) + \tau^{-1} l_k^n (z-1) (\kappa \tau + \nu \tau (2 l_k -1) + \ln 2) + (1 + z - 2 l_k z) (\kappa + \nu + \ln 2 / \tau)}{ 2 \nu (l_k - 1) + \tau^{-1} l_k^n (z-1) (2 l_k - 1) ( \kappa \tau + \nu \tau + \ln 2 ) + (1 + z - 2 l_k z) (\kappa + \nu + \ln 2 / \tau)} \right)^{m_0},
\end{equation}

where $l_k \equiv e^{\kappa \tau}$. Extinction probability is thus

\begin{equation}
P_{BD}(m=0) = \left( \frac{(l_k ^n - 1) (\kappa \tau + \nu \tau (2 l_k - 1) + \ln 2) }{\kappa \tau (2 l_k^{n+1} - l_k^n - 1) + \nu \tau (2 l_k -1)(l_k^n - 1) - \ln 2 - l_k^n \ln 2 + l_k^{n+1} \ln 4 } \right)^{m_0}
\end{equation}

which reduces to

\begin{equation}
P_{BD}(m = 0) \left( \frac{1 - l_k^n}{1 + l_k^n - 2l_k^{n+1} } \right)^{m_0}
\end{equation}

for $\nu = 0$, as given in the Main Text.

In the absence of cell divisions, the birth-death generating function is simply Eqn. \ref{nodivsolnsi} with $\alpha = 0$. Setting $z = 0$ gives the general extinction probability

\begin{equation}
P_{BD'}(m=0) = \left( \frac{ \nu e^{(\lambda - \nu) t} - \nu }{\lambda e^{(\lambda - \nu) t} - \nu} \right)^{m_0}.
\end{equation}

Copy number balance can be enforced in the absence of cell divisions by taking the $\lambda = \nu$ limit, from which follows the generating function

\begin{equation}
G_{BD',H}(z,t) = \left( \frac{ z + \nu t - \nu z t}{1 + \nu t - \nu z t} \right)^{m_0}
\end{equation}


from which straightforwardly follows the extinction probability

\begin{equation}
P_{BD', H}(m=0) = \left( \frac{\nu t}{1 + \nu t}\right)^{m_0}
\end{equation}

\subsection*{Other partitioning regimes}

We have derived results for the case where a binomially-distributed random number of agents is lost at each cell division. We now consider the case where this number is a fixed constant. We will denote this constant loss number by $\eta$. In this case, 

\begin{eqnarray}
P_{\delta}(m_{i,a} | m_{i,b}) & = & \delta_{m_{i,a}, m_{i,b} - \eta}; \\
\sum_{m_{i,a} = 0}^{m_{i,b}}  & \xi(z,t) & \left[ g(z,t) \right]^{m_{i, a}} P_{\delta}(m_{i,a} | m_{i,b}) \nonumber \\
& = &  \xi(z,t) g(z,t)^{-\eta} g(z,t)^{m_{i,b}}; \\
\text{and so}\,\, \phi(z,t) & = & g(z,t)^{-\eta}; \label{phicase1} \\
\theta(g(z,t)) & = & g(z,t).
\end{eqnarray}

As $\theta$, $\xi$ and $g$ take the same form as for the random loss case, the solutions for $z_i$ and $h_i$ are the same as before. The difference (due to the different form of $\phi$) is in the second product in Eqn. \ref{generalgf}, which is now $\prod_{i=1}^n g(z_i,\tau)^{-\eta}$. In the Appendix we show that this factor takes the form of Eqn. \ref{firstproductform} with $A_1 = \nu l (z-1) (-x_2)^n, B_1 = B_2 = x_1^n (\nu - \lambda z), A_2 = \lambda l (z-1) (-x_2)^n, \rho_A = x_1, \rho_B = (-x_2), \gamma = -\eta$.

The generating function in the case of deterministic subtractive inheritance involves (i) Eqn. \ref{siproductform1} applied to the appropriate terms described in Eqn. \ref{zilinksub}; (ii) Eqn. \ref{siproductform1} applied to Eqn. \ref{thirdproduct1} as described; (iii) Eqn. \ref{xisi}; and (iv) $h_0$ from Eqn. \ref{subtractivehsolnsi}, and is

\begin{eqnarray}
G_n(z,t) & = & \left( \frac{ (\lambda l' (z-1) (-x_2)^n + x_1^n (\nu - \lambda z)) \left( \frac{ -\nu l' (z-1)(-x_2)^n }{x_1^n (\nu - \lambda z) } ; \frac{x_1}{-x_2} \right)_{n+1} } { (\nu l' (z-1) (-x_2)^n + x_1^n (\nu - \lambda z) ) \left( \frac{-\lambda l' (z-1)(-x_2)^n}{x_1^n (\nu - \lambda z)} ; \frac{x_1}{-x_2} \right)_{n+1} } \right)^{- \eta} \nonumber \\
&& \times \left( \frac{ (\lambda l' l(z-1) + l (\nu - \lambda z)) \left( \frac{-\lambda l'(z-1)}{l(\nu - \lambda z)} ; l \right)_{n+1} }{ (\lambda l'(z-1) + l (\nu - \lambda z)) \left( \frac{ -\lambda l' l (z-1) }{l(\nu - \lambda z)} ; l \right)_{n+1} } \right)^{\alpha / \lambda} \nonumber \\
&& \times \left( \frac{\nu - \lambda}{\lambda l (z-1) - \lambda z + \nu } \right)^{\alpha / \lambda} \left( \frac{l^n l' \nu (z-1) + \nu - \lambda z}{l^n l' \lambda (z-1) + \nu - \lambda z} \right)^{m_0} 
\end{eqnarray}

Now we briefly explore two other inheritance regimes of potential biological applicability. In these cases we have not been able to obtain closed-form solutions for an arbitrarily large number of cell divisions: however, the appropriate recursion relations may be followed for as many divisions as required in order to obtain a closed-form solution for the generating function.

First we consider deterministic partitioning of agents, where each daughter inherits exactly half of a parent's population. In this case:

\begin{eqnarray}
P_{\delta}(m_{i,a} | m_{i,b}) & = & \delta_{m_{i,a},m_{i,b}/2}; \\
\sum_{m_{i,a} = 0}^{m_{i,b}}  & \xi(z,t) & \left[ g(z,t) \right]^{m_{i, a}} P_{\delta}(m_{i,a} | m_{i,b}) \nonumber \\
& = &  \xi(z,t) g(z,t)^{m_{i,b} / 2}; \\
\text{and so}\,\, \phi(z,t) & = &  1; \\
\theta(g(z,t)) & = & \sqrt{g(z,t)},
\end{eqnarray}

leading to the recurrence relations

\begin{eqnarray}
z_i & = & \sqrt{g(z_{i-1}, \tau)} \,;\, z_1 = \sqrt{g(z,t)}. \label{determrecur1}\\
h_i & = & g \left( \sqrt{h_{i+1}}, \tau \right) \,;\, h_n = g (z, t). \label{determrecur2}
\end{eqnarray}

Next we consider the binomial inheritance of clusters of agents. We will assume that these clusters are of fixed size $n_c$. In this case, we consider the new variables $C_b = m_{i,b} / n_c$, $C_a = m_{i,a} / n_c$ (denoting the number of clusters before and after a cell division), and write

\begin{eqnarray}
&& \sum_{m_{i,b} = 0}^{\infty} \sum_{m_{i,a} = 0}^{m_{i,b}} g^{m_{i,a}} P_{\delta}(m_{i,a} | m_{i,b}) P_{i-1}(m_{i,b}, \tau | m_0) \nonumber \\
& = & \sum_{C_b=0}^{\infty} \sum_{C_a=0}^{C_b} g^{n_c C_a} \binom{C_b}{C_a} 2^{-C_b} P_{i-1}(n_c C_b, \tau | m_0) \\
& = & \sum_{C_b=0}^{\infty} \left( \frac{1}{2} + \frac{g^{n_c}}{2} \right)^{C_b} P_{i-1}(n_c C_b, \tau | m_0) \\
& = & \sum_{m_{i,b} = 0}^{\infty} \left( \frac{1}{2} + \frac{g^{n_c}}{2} \right)^{\frac{m_{i,b}}{n_c}} P_{i-1}(m_{i,b}, \tau | m_0)
\end{eqnarray}

The resultant generating function analysis yields a very similar outcome to that in Eqns. \ref{induct1}-\ref{induct2}, with the altered recurrence relation

\begin{eqnarray}
z_i & = & \left( \frac{1}{2} + \frac{g(z_{i-1}, \tau)^{n_c}}{2} \right)^{\frac{1}{n_c}} \,;\, z_1 = \left( \frac{1}{2} + \frac{g(z,t)^{n_c}}{2} \right)^{\frac{1}{n_c}} \label{clusrecur1} \\
h_i & = & g_0 \left( \left( \frac{1}{2} + \frac{h_{i+1}^{n_c}}{2} \right)^{\frac{1}{n_c}}, \tau \right) ; h_n = g_0(z,t). \label{clusrecur2}
\end{eqnarray}

We have been unable to reduce the recurrence relations Eqns. \ref{determrecur1}-\ref{determrecur2} or Eqns. \ref{clusrecur1}-\ref{clusrecur2} to a closed-form solution for birth-death dynamics, but the corresponding problems may be solved for an arbitrary number of cell divisions by writing out the recurrence explicitly, thereby obtaining the generating function for a given number of cell divisions. The figure in the main text uses this approach.

\end{document}